\documentclass[useAMS,usenatbib]{mn2e}
\usepackage[T1]{fontenc}
\usepackage{aecompl}
\usepackage{amsmath}
\usepackage{multicol}
\usepackage{epsfig}
\usepackage{rotating}
\def\USydney{$^{1}$}
\def\CAASTRO{$^{2}$}
\def\IT{$^{3}$}
\def\UWisc{$^{4}$}
\def\CSIRO{$^{5}$}
\def\Curtin{$^{6}$}
\def\Monash{$^{7}$}
\def\SKAalt{$^{8}$}
\def\SKA{$^{9}$}
\def\CfA{$^{10}$}
\def\ASU{$^{11}$}
\def\ANU{$^{12}$}
\def\Haystack{$^{13}$}
\def\RRI{$^{14}$}
\def\MIT{$^{15}$}
\def\UW{$^{16}$}
\def\Victoria{$^{17}$}
\def\UMelbourne{$^{18}$}
\def\Tata{$^{19}$}
\def\UTasmania{$^{20}$}

\title[M. E. Bell et al]{A survey for transients and variables with the Murchison Widefield Array 32-tile prototype at 154 MHz}
\author[M.E.Bell et al]{M. E. Bell\USydney$^{,}$\CAASTRO,
T. Murphy\USydney$^{,}$\CAASTRO$^{,}$\IT,
D. L. Kaplan\UWisc,
P. Hancock\USydney$^{,}$\CAASTRO,
B. M. Gaensler\USydney$^,$\CAASTRO
\newauthor
J. Banyer\USydney,
K. Bannister\CSIRO$^{,}$\CAASTRO,
C. Trott\Curtin$^,$\CAASTRO,
N. Hurley-Walker\Curtin,
R.~B.~Wayth\Curtin$^,$\CAASTRO, 
\newauthor
J.~-P.~Macquart\Curtin$^,$\CAASTRO,
W.~Arcus\Curtin, 
D.~Barnes\Monash,
G.~Bernardi\SKAalt$^,$\SKA,
J.~D.~Bowman\ASU, 
\newauthor 
F.~Briggs\ANU$^{,}$\CAASTRO,
J.~D.~Bunton\CSIRO, 
R.~J.~Cappallo\Haystack,
B.~E.~Corey\Haystack,
A.~Deshpande\RRI,
\newauthor
L.~deSouza\CSIRO$^,$\USydney,
D.~Emrich\Curtin,
R.~Goeke\MIT,
L.~J.~Greenhill\CfA,
B.~J.~Hazelton\UW, 
D.~Herne\Curtin,
\newauthor 
J.~N.~Hewitt\MIT, 
M.~Johnston-Hollitt\Victoria,
J.~C.~Kasper\CfA, 
B.~B.~Kincaid\Haystack, 
R.~Koenig\CSIRO, 
\newauthor
E.~Kratzenberg\Haystack ,
C.~J.~Lonsdale\Haystack, 
M.~J.~Lynch\Curtin, 
S.~R.~McWhirter\Haystack,
D.~A.~Mitchell\UMelbourne$^,$\CAASTRO, 
\newauthor
M.~F.~Morales\UW,
E.~Morgan\MIT,  
D.~Oberoi\Tata, 
S.~M.~Ord\Curtin$^,$\CAASTRO,
J.~Pathikulangara\CSIRO, 
T.~Prabu\RRI, 
\newauthor
R.~A.~Remillard\MIT, 
A.~E.~E.~Rogers\Haystack, 
A.~Roshi\RRI, 
J.~E.~Salah\Haystack, 
R.~J.~Sault\UMelbourne, 
\newauthor
N.~Udaya~Shankar\RRI, 
K.~S.~Srivani\RRI, 
J.~Stevens\CSIRO$^,$\UTasmania, 
R.~Subrahmanyan\RRI$^,$\CAASTRO, 
S.~J.~Tingay\Curtin$^,$\CAASTRO, 
\newauthor
M.~Waterson\Curtin$^,$\ANU,
R.~L.~Webster\UMelbourne$^,$\CAASTRO, 
A.~R.~Whitney\Haystack, 
A.~Williams\Curtin,
C.~L.~Williams\MIT,
\newauthor
J.~S.~B.~Wyithe\UMelbourne$^,$\CAASTRO\\
$^{1}$Sydney Institute for Astronomy, School of Physics, The University of Sydney, NSW 2006, Australia\\
$^{2}$ARC Centre of Excellence for All-sky Astrophysics (CAASTRO)\\
$^{3}$School of Information Technologies, The University of Sydney, NSW 2006, Australia\\
$^{4}$University of Wisconsin--Milwaukee, Milwaukee, USA\\
$^{5}$CSIRO Astronomy and Space Science, Australia\\
$^{6}$International Centre for Radio Astronomy Research, Curtin University, Perth, Australia\\
$^{7}$Monash e-Research Centre, Monash University, Melbourne, Australia \\
$^{8}$Department of Physics and Electronics, Rhodes University, PO Box 94, Grahamstown, 6140, South Africa\\
$^{9}$SKA SA, 3rd Floor, The Park, Park Road, Pinelands, 7405, South Africa\\
$^{10}$Harvard-Smithsonian Center for Astrophysics, Cambridge, USA\\
$^{11}$Arizona State University, Tempe, USA\\
$^{12}$The Australian National University, Canberra, Australia\\
$^{13}$MIT Haystack Observatory, Westford, USA\\
$^{14}$Raman Research Institute, Bangalore, India\\
$^{15}$MIT Kavli Institute for Astrophysics and Space Research, Cambridge, USA\\
$^{16}$University of Washington, Seattle, USA\\
$^{17}$Victoria University of Wellington, New Zealand\\
$^{18}$The University of Melbourne, Melbourne, Australia\\
$^{19}$Tata Institute for Fundamental Research, Pune, India\\
$^{20}$University of Tasmania, Hobart, Australia\\
}

\begin{document}
%\date{Accepted xxxx December xx. Received xxxx December xx; in original form xxxx October xx}
\pagerange{\pageref{firstpage}--\pageref{lastpage}} \pubyear{2012}
\maketitle
\label{firstpage}
\begin{abstract}
We present a search for transient and variable radio sources at 154 MHz with the Murchison Widefield Array 32-tile prototype. Fifty-one images were obtained that cover a field of view of $ 1430$ deg$^{2}$ centred on Hydra A. The observations were obtained over three days in 2010 March and three days in 2011 April and May. The mean cadence of the observations was 26 minutes and there was additional temporal information on day and year timescales. We explore the variability of a sample of 105 low frequency radio sources within the field. Four bright ($S>6$~Jy) candidate variable radio sources were identified that displayed low levels of short timescale variability (26 minutes). We conclude that this variability is likely caused by  simplifications in the calibration strategy or ionospheric effects. On the timescale of one year we find two sources that show significant variability. We attribute this variability to either refractive scintillation or intrinsic variability. No radio transients were identified and we place an upper limit on the surface density of sources $\rho<7.5 \times 10^{-5}$ deg$^{-2}$ with flux densities $>5.5$~Jy, and characteristic timescales of both 26 minutes and one year.    
\end{abstract}
\begin{keywords}
instrumentation: interferometers, radio continuum: general, techniques: image processing, catalogues
\end{keywords}

%%%%%%%%%%%%%%%%%%%%%%%%%%
\section{Introduction}
%%%%%%%%%%%%%%%%%%%%%%%%%%

Historically, low frequency variability studies (300~MHz to 1~GHz) have focused on quantifying the properties of discrete samples of bright extragalactic radio sources (e.g. \citealt{Hunstead_72}; \citealt{Cotton76a}, b;  \citealt{Condon_79}; \citealt{Dennison_81}; \citealt{Fanti81}; \citealt{Gregorini}; \citealt{Spangler_1989}; \citealt{Riley1995}; \citealt{Riley_1998}; \citealt{Gaensler_2000}). Targeted surveys like these do not blindly search for or quantify the abundance of variables and transients.    

A small number of very low frequency surveys ($<$200 MHz) have performed blind searches for variable radio sources. \cite{Mcgilchist_1990} used two epochs of data separated by approximately one year, to investigate the variability of 811 unresolved extragalactic radio sources at 151~MHz (also see \citealt{Riley_1993} and \citealt{Minns2000}). No radio sources with a fractional variability $>$4\% were found at flux densities $>$3~Jy. 

\cite{Mcgilchist_1990} conclude that variability on timescales of one year is extremely rare at 151 MHz. In contrast, \cite{slee_1988} report on the variability of a sample of 412 extragalactic radio sources with $S>$1~Jy at 80 and 160~MHz using the Culgoora Circular Array \citep{Culgoora_Telescope}. Within this sample, more than 23\% of the sources show fractional flux density changes greater than $>$4\% at 160~MHz (discussed by \citealt{Mcgilchist_1990}). The disparity between these results can potentially be accounted for by the different properties of the instruments and techniques used to obtain the samples. 
 
In the surveys described above (with the exception of \citealt{slee_1988}) most source are found to be non-variable, with about $1\%$ of sources showing variability. The focus for future low frequency instruments and surveys is for complete and unbiased blind surveys for both variables and transients, and to be able to respond in real time to such events with coordinated multi-wavelength followup. Timely followup is critical for transient classification and the lack of immediate follow-up has impeded progress within the field. 

The two most significant recent results in this area are a 74~MHz blind survey by \cite{Lazio2011} and a 325~MHz survey by
\cite{Jaeger2012}. 
\citet{Lazio2011} conducted a survey for transients using the Long Wavelength Demonstrator Array (LWDA). No transients were reported above a 5$\sigma$ threshold of 2500 Jy over an area of $\sim$10,000 deg$^{2}$. At the other extreme of flux density, sensitivity and field of view, \citet{Jaeger2012} conducted the deepest blind transients survey to date below 500~MHz. Six epochs of Very Large Array (VLA) observations centred on the Spitzer-space-telescope Wide-field InfraRed Extragalactic (SWIRE) deep field were searched for transient and variable radio sources above 2.1~mJy (10$\sigma$) at 325~MHz. One radio transient was reported in 72 hours of observing, implying a surface density of transients over the whole sky of 0.12~deg$^{-2}$. 

The Galactic centre has been the focus of low frequency radio transient surveys and a number of detections have been made. \cite{Hyman_Nature} report the detection of a bursting and periodic coherent radio source with peak flux density $\sim$1~Jy (also see Hyman et al. 2006). The serendipitous detection of a steep spectrum radio transient with peak flux of 100~mJy at 235 MHz was reported by Hyman et al. (2009), also see Hyman et al. (2002). 

At low frequencies, a number of interferometers are (or soon will be) utilising dipole based technologies to produce sensitive, high resolution wide field instruments. These telescopes include: the Low Frequency Array (LOFAR; \citealt{LOFAR_paper}), the Long Wavelength Array (LWA; \citealt{LWA}), the Murchison Widefield Array (MWA; \citealt{MWA_lonsdale}; \citealt{Tingay2012}; \citealt{Bowman2012}) and potentially the low frequency component of the Square Kilometre Array (SKA; \citealt{SKA_paper}). These wide field telescopes will enable all-sky monitor (and alert) type functionality for the detection of radio transients and variables in both the Northern and Southern Hemispheres. 

The MWA is a low frequency radio interferometer and SKA precursor instrument. It consists of 128 tiles (each containing 16 dipoles) operating in the frequency range 80$-$300~MHz. A 32-tile prototype instrument (which we will refer to as the MWA-32T) was operated on the site until mid-2011 (see \citealt{Ord2010}). In this paper we use the MWA-32T to explore the temporal behaviour of a sample of bright low frequency sources in an area of sky 1430~deg$^2$ centred on the bright radio galaxy Hydra~A (3C 218). 

A number of successful surveys have so far been carried out with the MWA-32T (i.e. \citealt{Bernardi}). In particular, \cite{WIlliams2012} used multiple MWA-32T snapshot observations obtained in 2010 of the field centred around Hydra A to produce a single deep broadband image (110 to 200 MHz) with $>$2500~deg$^{2}$ field of view. A catalogue of over 655 sources was presented and compared with existing radio catalogues. In this paper we use snapshot observations at 154 MHz of the field centred on Hydra~A from both 2010 and 2011 to search for transient and variable radio sources. 
We restrict the analysis to data centred at 154 MHz, with a bandwidth of 30.72 MHz; we also image a slightly smaller field of view.

In Section 2 of this paper we present a description of the MWA-32T instrument and our observations. 
In Section 3 we present the data reduction procedures including  calibration, imaging and primary beam correction of the data. 
In Section 4 we present the Variables and Slow Transients (VAST; \citealt{VASTPaper}) prototype pipeline. 
We use a sequence of images in Section 5 to demonstrate the science quality of the observations including: flux stability, positional accuracy and catalog cross-matched flux density comparisons. 
In Section 6 we discuss the transient and variability search results and we place constraints on the prevalence of dynamic radio phenomena at these survey frequencies, flux densities and cadences.  

%%%%%%%%%%%%%%%%%%%%%%%%%%
\section{MWA-32T Observations}
%%%%%%%%%%%%%%%%%%%%%%%%%%
The MWA-32T was a 32-tile engineering prototype operated at the Murchison Radio Observatory (MRO). The telescope was decommissioned in late 2011 to be replaced by the full MWA. The purpose of the MWA-32T was to test the essential technologies and software pipelines required for the deployment of the MWA. The 32-tiles were arranged in an approximately circular configuration with a pseudo-random spacing of tiles around the area. The array had a maximum baseline length of $\sim 350$m, this dense configuration provided excellent snapshot $uv$ coverage \citep{Ord2010} at a resolution of 15$^{\prime}$ at 154 MHz. 

Each tile consisted of 16 dual-polarisation dipoles co-located on a metal ground screen and were sensitive to a frequency range 80$-$300 MHz. An analogue beam former was used to insert phase delays into the signal path to steer the beam around the sky. One of the most important features of both the MWA-32T and the full MWA telescope for transient science is the field of view: a full-width half-maximum of $\sim$25$^{\circ} / (\nu/150 $MHz) can be realised at zenith. For a more detailed description of the antenna design, instrument and science drivers see \cite{Bowman2007}, \cite{MWA_lonsdale}, \cite{Tingay2012} and \cite{Bowman2012}.  

The observations presented in this paper were obtained in 2010 March and in 2011 April and May. A total of 51 observations were used in our analysis. Table~\ref{obs_table} summarises the dates of the observations. In both 2010 and 2011 the observations were spread over three different days. Each observation was five minutes in length and a total of 4.25 hours of data were obtained. 

\begin{figure}
\centering
\includegraphics[scale=0.90]{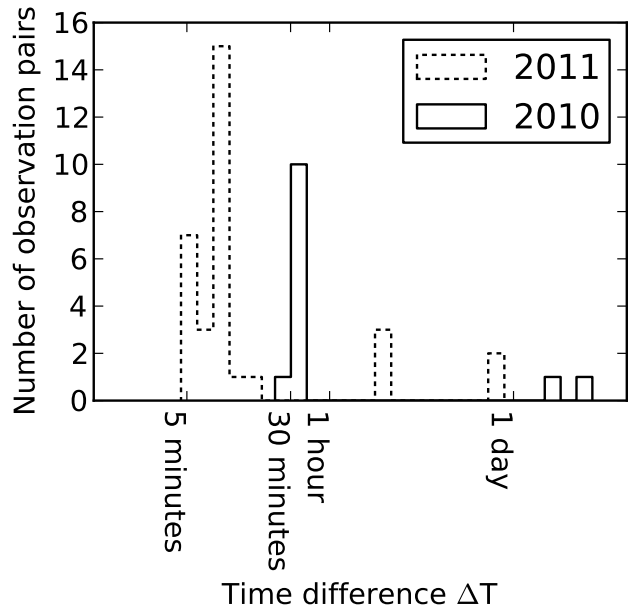}
\caption{A histogram of the (log) time difference $\Delta T$ between consecutive observation pairs.}
\label{hist}
\end{figure}

Figure~\ref{hist} shows a histogram of the time difference between consecutive observation pairs (excluding the year gap between 2010 and 2011). The observations in 2010 typically sampled timescales of $\sim$30 minutes, with two separate epochs that were separated by two and four days respectively. The 2011 observations were typically spaced closer together in time with a dominant timescale of $\sim$10 minutes; two epochs were both spaced by one day. By comparing the data obtained in both 2010 and 2011 we can also assess temporal changes on the timescale of approximately one year. 

The observations were centred close to the source Hydra~A (3C 218) at RA = 9$^{h}$18$^{m}$6$^{s}$ and $\delta$ = $-$12$^{\circ}$5$^{\prime}$45$^{\prime \prime}$  (J2000). Hydra~A is a Fanaroff-Riley type I (FR-I) radio Galaxy at a redshift $z=0.0542$ \citep{Taylor1996} with large scale radio structure extending a total angular extent of  8$^{\prime}$ \citep{Lane2004}. It is bright and almost unresolved on MWA-32T baselines, making it an ideal in-field calibrator. 
  
\begin{table}
\centering
\caption{Summary of observations at 154 MHz. A total of 51 observations (4.25 hours of data) were used for the analysis presented in this paper.}
\begin{tabular}{|r|r|r|r|r|}
\hline
\multicolumn{1}{c}{Date} & \multicolumn{1}{c}{N. of obs.} & \multicolumn{1}{c}{Hour angle range (hours)} & \\
\hline
2010/03/22 & 7 & $-1.36$~to~2.13\\
2010/03/26 & 5 & $-1.68$~to~0.44\\
2010/03/28 & 3 & ~~$-1.55$~to~$-0.38$\\
\hline
2011/04/29 & 10 & $-0.81$ to 3.06\\
2011/04/30 & 12 & $-0.78$ to 2.92\\
2011/05/01 & 14 & $-1.61$ to 3.01\\
\hline
\label{obs_table}
\end{tabular}
\end{table}

%%%%%%%%%%%%%%%%%%%%%%%%%%
\section{Data reduction}
%%%%%%%%%%%%%%%%%%%%%%%%%%

The full MWA will use a custom real-time system (RTS) to flag, calibrate and image the correlated visibilities being produced by the telescope (see \citealt{Mitch_2008} for an overview). However, for our analysis we used a standard imaging pipeline, developed by \cite{WIlliams2012} and written in the {\sc casa}\footnote{http://casa.nrao.edu} environment.  
A small number of adjustments were made to promote compatibility between this imaging pipeline and the VAST pipeline. We
discuss the notable parts of the imaging process below. For further discussion see \cite{WIlliams2012}.   

%%%%%%%%%%%%%%%%%%%%%%%%%%
\subsection{Pre-processing} 
%%%%%%%%%%%%%%%%%%%%%%%%%%
Fixed beamformer settings were typically used to steer the telescope pointing around the sky in quantised steps. For these observations, Hydra A was not always at the pointing centre or phase centre of the instrument (see Figure \ref{X15_image}). To enable accurate calibration, phase shifts were applied to all visibilities such that Hydra A was positioned at the phase centre. These data were then averaged from one-second sampling down to four seconds with 40-kHz spectral resolution. Finally these data were flagged for radio frequency interference (RFI) and then stored in UVFITS format.

%%%%%%%%%%%%%%%%%%%%%%%%%%
\subsection{Flagging and calibration} 
%%%%%%%%%%%%%%%%%%%%%%%%%%
\label{cal}
The UVFITS files were converted to measurement-set format and imported into the {\sc casa} (version 3.4) environment. Known problematic channels were flagged out before proceeding. The visibilities were then further flagged using the {\sc casa} task {\sc testautoflag} (with a time window of four samples and a frequency window of three channels). Typically a few percent of all visibilities were flagged due to RFI.  The shortest baselines ($<50 \lambda$) were flagged out to suppress diffuse Galactic emission contaminating accurate calibration. The flux density of Hydra A  was set using the task {\sc setjy} using a reference flux density taken from the VLA Low-frequency Sky Survey  (VLSS; \citealt{VLSS}) catalogue ($S_{VLSS} = 579.6$ Jy at 74 MHz with a spectral index $\alpha=-0.89$). Extrapolating the flux density of Hydra~A at 74~MHz to 154~MHz gives $S_{MWA}=301.5$~Jy. 

The visibilities were calibrated using the {\sc casa} task {\sc bandpass}, which generated time-independent, frequency-dependent antenna gain solutions. Hydra~A dominated the visibilities within this field and a simple point source model was used for flux and phase calibration. Examining a $uv$ distance versus amplitude plot (post-calibration) showed that Hydra~A was slightly resolved at these frequencies. We consider this a potential source of error, to be considered in the analysis presented in Section \ref{results}. More complex sky models will be required for future high resolution calibration and imaging. Post calibration flagging was applied (with the same time and frequency windows defined above) to remove badly calibrated and erroneous visibilities. Before imaging, the visibilities were split into four subbands centred at 143, 150, 158 and 166 MHz (each with 7.68 MHz bandwidth). 

%%%%%%%%%%%%%%%%%%%%%%%%%%
\subsection{Imaging} 
%%%%%%%%%%%%%%%%%%%%%%%%%%
The calibrated visibilities for all the subbands and full bandwidth data were imaged using the {\sc casa} task {\sc clean}. The XX and YY polarisations were imaged separately because a different instrumental beam correction is required per polarisation. Images were created with an image size of 1024$\times$1024 pixels and {\sc clean}ed with up to 2000 iterations using the Cotton-Schwab algorithm (see \citealt{Schwab}) down to a limiting threshold of 300 mJy. Different {\sc clean} scenarios were tested (via experimentation with iterations, thresholds, algorithm etc) that optimised the balance between final image RMS and false positives caused by over {\sc clean}ing. A cell size of 2.5$^{\prime}$ was used and yielded a total field of view of 42.7$^{\circ}$ $\times$ 42.7$^{\circ}$. A uniform weighting scheme was utilised for imaging. The w-projection algorithm \citep{Cornwall} was used to correct for wide field effects.   

\begin{figure*}
\centering
\begin{tabular}{cc}
\hspace{-50pt}
\includegraphics[scale=0.63]{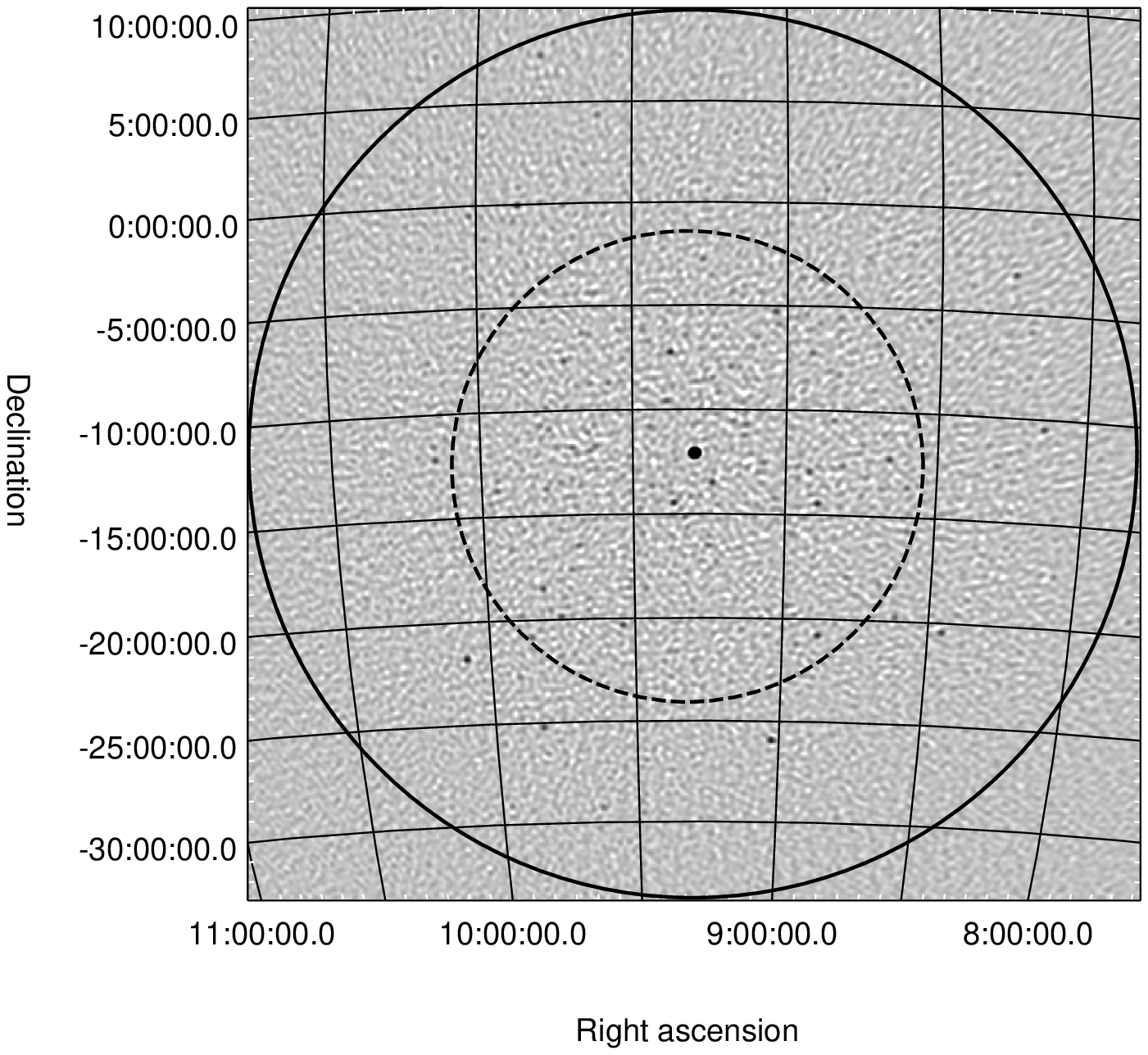}
\hspace{-80pt}
\includegraphics[scale=0.63]{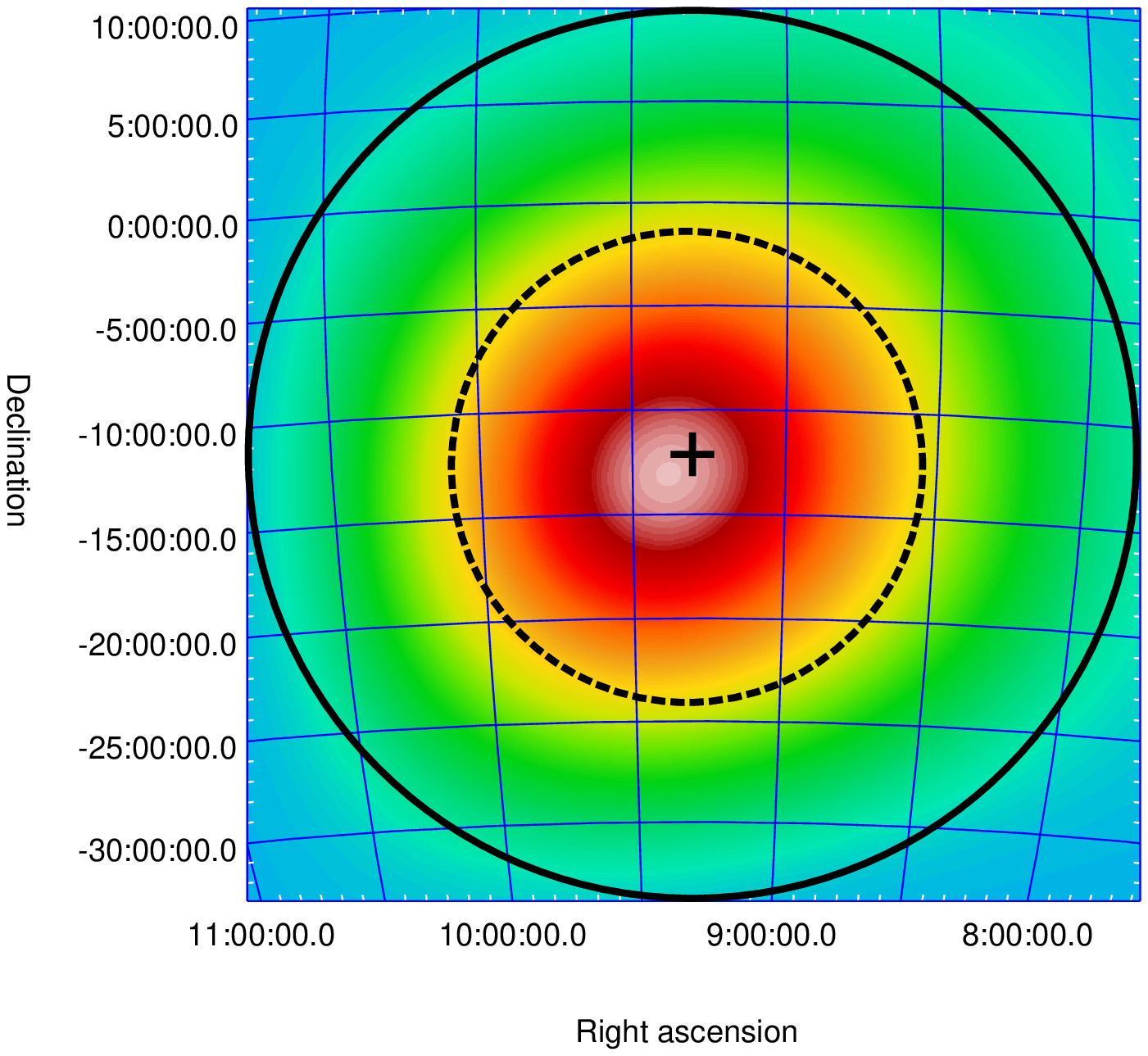}
\end{tabular}
\caption{Left panel: A five minute snapshot observation of Hydra A taken on 2010-03-22 (XX polarisation). The grey scales varies logarithmically from -3 to 300 Jy~beam$^{-1}$. This observation had an RMS of 369 mJy beam$^{-1}$ (without primary beam correction) and beam size $15.5^{\prime} \times 13.2^{\prime}$. Right panel: The corresponding primary beam map for this observation. The location of Hydra~A is indicated with a black cross. The primary map is normalised to one and the colour scale runs from 0.1 (blue) to one (white). In this example the peak primary beam location is not centred on Hydra~A. The dashed line denotes the approximate region of half-power and the solid line denotes the solid angle that is considered in the analysis presented in this paper (radius$ = 21.35$ degrees).}
\label{X15_image}
\end{figure*}

Figure \ref{X15_image} shows an example XX polarisation image produced by the imaging pipeline (observation date 2010-03-22). The corresponding primary beam map (discussed in Section \ref{PB}) is also shown. The beam size of this images is $15.5^{\prime} \times 13.2^{\prime}$. The dashed circular line shows the approximate location of the half-power radius and the solid black line shows the region used for the analysis presented in this paper (radius$ = 21.35$ degrees). The 2010-03-22 observation had an RMS of 369~mJy~beam$^{-1}$ (without primary beam correction). This value was calculated using an interquartile range that excluded bright sources from the calculation. \cite{WIlliams2012} reported that the stacked MWA-32T observations of Hydra~A had a confusion limited noise of $\sim$160 mJy.   

%%%%%%%%%%%%%%%%%%%%%%%%%%
\subsection{Primary beam correction} 
%%%%%%%%%%%%%%%%%%%%%%%%%%
\label{PB}
We used the method and algorithm presented by \cite{WIlliams2012} and \cite{Mitch_2012} to produce primary beam maps for each polarisation and frequency. Figure~\ref{X15_image} shows an example primary beam map for the observation 2012-03-22. This map is normalised to one at the peak response of the instrument and the location of the peak response is a function of time, polarisation and frequency. The primary beam maps give a monochromatic solution across the full 30.72 MHz bandwidth (compared to 7.68 MHz for the subbands), which may be a potential source of error. The primary beam maps for each of the subbands provide a more accurate frequency dependent correction.  

The Stokes~I images were produced by firstly multiplying the XX and YY images by the XX and YY polarisation primary beam maps. The corrected XX and YY maps were then averaged together to form the Stokes~I images. It should be noted that there was a certain amount of correlation between the off source pixel properties of the XX and YY polarisation images, because both the XX and YY images were dominated by confusion noise. This meant that per-snapshot, both the XX and YY dipoles see the same flux on the sky, in the same position, at the same time, with the same synthesised beam. By forming the Stokes~I images, the correlation between the XX and YY images was successfully accounted for in the resulting RMS budget of the Stokes I images. The noise properties of the images were used to place errors on the source fluxes, and the fluxes and associated errors were used in the subsequent variability analysis. It was therefore critical to use the correct flux errors calculated from the Stokes~I images  in the analysis. 

To search the 2010 and 2011 images for long term transient and variable behaviour, we averaged all Stokes~I images from 2010 and 2011 to form ``deep" images (as done in \citealt{WIlliams2012}). Each of the snapshots used in the average had slightly different beam properties. To account for this we fitted an elliptical Gaussian at the position of Hydra~A in both the 2010 and 2011 averaged images. We then used the resulting fitting properties of Hydra A (major/minor axis size and position angle) as the overall beam values for the images.    

%%%%%%%%%%%%%%%%%%%%%%%%%%
\section{The Variables and Slow Transients Detection Pipeline}
%%%%%%%%%%%%%%%%%%%%%%%%%%
A prototype transients and variables detection pipeline has been developed which is used for the analysis presented in this paper (see \citealt{Banyer2012} and \citealt{VASTPaper}). The pipeline works as follows:
(i) For each Stokes~I (full bandwidth) image, a source extraction algorithm was used to locate and characterise sources. In the subband images, elliptical Gaussian fits were calculated at the locations of the sources found in the full bandwidth image. The results from the source extractor including all metadata were stored in a PostgreSQL\footnote{http://www.postgresql.org/} SQL database.  
(ii) All extracted sources were associated positionally (through the full bandwidth image and subbands). 
(iii) All source positions were ``monitored", i.e. if a source detected in one image was not found in another image, a measurement (described in Section \ref{monitoring}) was taken at that position and stored in the database. 
(vi)~The light-curves for all sources were tested for significant variability or transient behaviour. 
(vii) The sources were crossmatched with multi-wavelength catalogues, and (radio) spectral energy distribution (SED) plots were created. Finally, the results (light-curves, SEDs, source thumbnail images, variability statistics etc) were viewed via a web interface. % (based on the Django\footnote{https://www.djangoproject.com/} platform). 

%%%%%%%%%%%%%%%%%%%%%%%%%%
\subsection{Source extraction}
%%%%%%%%%%%%%%%%%%%%%%%%%%
The {\sc aegean} source extraction algorithm (version 808) was used for the analysis presented in this paper \citep{Paul_AEGEAN}. The source finding stage of the VAST pipeline results in a set of measurements of the position (RA,Dec), flux (peak and integrated), and shape (major/minor axis and position angle) of each source, as well as their associated uncertainties. We use the peak flux from the fitting procedure in all subsequent variability analysis. Further details of the {\sc aegean} source finding algorithm can be found in \cite{Paul_AEGEAN}.

We used a background size of 150$\times$150 pixels which corresponds to $\sim$78.5 synthesised beam areas (assuming a circular beam of width five pixels). A source extraction level of $5.5\sigma$ was used, which corresponds to a false detection rate of 2$\times 10^{-3}$ per synthesised beam, or 0.1 false detections for every 51 images. 

%%%%%%%%%%%%%%%%%%%%%%%%%%
\subsection{Monitoring}
%%%%%%%%%%%%%%%%%%%%%%%%%%
\label{monitoring}
If a source was not detected by {\sc aegean} above the source extraction threshold a ``forced" measurement was performed. This was done by assuming a point source morphology at the best known location of the source. The flux measuring process was identical to a fitting routine where the morphology and position of the source were held fixed and only the flux was allowed to vary. The measured flux and associated error were obtained directly using the image data, $I_{ij}$, and a model beam, $B_{ij}$, scaled to unit flux. The flux, $S$, and associated error, $\sigma$, are then given by:

\begin{equation}
S = \frac{\sum_{i,j}I_{ij}}{\sum_{i,j}B_{ij}},
\end{equation}

\begin{equation}
\sigma = \sqrt{ \frac{\sum_{ij}B_{i,j}\left(\frac{I_{ij}}{B_{ij}}-S \right)^2}{\sum_{i,j}B_{i,j}} },
\end{equation}

\noindent where $N$ is the number of pixels enumerated by $i,j$. In the above calculation, we only consider pixels within the FWHM of the beam ($B_{ij}\geq 0.5$). For point sources that are above the blind-detection threshold, Equations 1 and 2 give flux measurements equal to that of {\sc aegean}, with error bars that are only slightly larger. By using forced measurements, we obviate the need to use upper limits from non-detections. This allows us to simplify the statistical interpretation and analysis because for each source in every image, we have a measurement of flux and error. 

\subsection{Source association} 
A source association algorithm was used to associate MWA sources as a function of time and frequency. From a sequence of images (in time) an association was defined as the closest source, within a synthesised beam diameter (typically $\sim 15^{\prime}$), to the average position of a given source formed from previous associations, as a function of frequency. 

We used the median RA and Dec of each source to associate them with external radio catalogues (the VLSS survey at 74 MHz (\citealt{VLSS}); the NRAO VLA Sky Survey at 1.4 GHz (NVSS; \citealt{Condon_NVSS}); the Parkes-MIT-NRAO survey at 4.8 GHz (PMN; \citealt{PMN}) and the Culgoora catalogue at 160 MHz (\citealt{Culgoora_CAT})).
A catalogue association was defined as the closest and brightest catalogue source to the median MWA source position. Only catalogue sources within one beam diameter of the MWA source positions were considered as crossmatches. 

%%%%%%%%%%%%%%%%%%%%%%%%%%
\subsection{Variability searches}
%%%%%%%%%%%%%%%%%%%%%%%%%%
\label{var_section}
We test for variability by calculating the $\chi^{2}$ probability that the source remained constant over the observing period (this method was also used by \citealt{Kesteven}; \citealt{Gaensler_2000} and \citealt{Keith2011}). For each source light-curve we calculate $\chi^{2}_{lc}$ where

\begin{equation}
\chi_{lc}^{2} = \sum_{i=1}^{n} \frac{(S_{i} - \tilde{S})^{2}}{\sigma_{i}^{2}}.
\label{chisquared}
\end{equation}

\noindent $S_{i}$ is the i$th$ flux density measurement with variance $\sigma^{2}_{i}$ within a source light-curve; $S_{i}$ and $\sigma_{i}$ are calculated as per Equations 1 and 2 respectively. $n$ is the total number of flux density measurements in the light-curve. $\tilde{S}$ is the weighted mean flux density defined as 

\begin{equation}
\tilde{S} = \sum_{i=1}^{n} \left( \frac{ S_{i}}{\sigma_{i}^{2}} \right) /  \sum_{i=1}^{n} \left( \frac{1}{\sigma_{i}^{2}} \right).
\label{w_mean}
\end{equation}

Assuming that the errors in source flux density measurements are drawn from a normal random distribution, and that the measurements are independent (see discussion in Section \ref{var_search}),  we would expect the values of $\chi^{2}_{lc}$ to follow the theoretical distribution $\chi^{2}_{T}$ (for $n-1$ degrees of freedom). For each light-curve we therefore calculate, using the cumulative distribution function (CDF), the probability $P$ that the $\chi^{2}_{lc}$ value could be produced by chance. We then define a variable source as having $P< 0.001$ and a non-varying source as having $P>0.001$. Note, the $P$-value is independent of the number of degrees of freedom so it provides a direct way to compare statistics from light-curves formed from different numbers of measurements. Furthermore, the $P$-value is used to search for sources based on the significance of the variability rather than the magnitude of the variability.  

In conjunction with the $\chi^{2}$ test we also calculated the modulation index and de-biased modulation index for all source light-curves.  The $\chi^{2}$ statistic is used first to test if a source is varying and then we use the modulation index and de-biased modulation index to quantify the degree of variability. We define the modulation index as   

\begin{equation}
m = \frac{\sigma_{S}}{\overline{S}} ,
\label{mod_index}
\end{equation}

\noindent where $\sigma_{S}$ is the standard deviation of the flux density of the source light-curve; and $\overline{S}$ is the mean flux density of the source light-curve. $\overline{S}$ in this case is the mean flux density, not the weighted mean flux density defined in Equation \ref{w_mean}. The modulation index is heavily dependent on the detection threshold of the source(s) in the images. For example, a non-varying source that had a mean detection of $\overline{S} = 5\sigma$ will yield a modulation index of $m = 20\%$ due to the statistical uncertainty in the measurement. The modulation index was still calculated because it was useful to compare with other variability studies reported in the literature. We also calculated the de-biased modulation index, which incorporates the errors on the flux measurements (see \citealt{Akritas1996}; \citealt{Barvainis}; \citealt{Sadler_2006}) and is defined as 

\begin{equation}
 m_{d} = \frac{1}{\overline{S}} \sqrt{\frac{\sum_{i=1}^{n}(S_{i} - \overline{S})^{2}   - \sum_{i=1}^{n} \sigma_{i}^{2}  }{n}} .
\label{debiased_mod_index}
\end{equation}

\noindent When $m_{d}$ is imaginary, we interpret this to mean the magnitude of the variability is uncertain, based on the size of the errors $\sigma_{i}$ (see \citealt{Barvainis}).  

Following \cite{Mcgilchist_1990}, we inspect the long term variability between the observations in 2010 and 2011, using the fractional variability, defined as

\begin{equation}
f = \frac{\Delta S}{\hat S}  = \frac{{S_{2010}} - {S_{2011}}}{\frac{1}{2}({S_{2010}} + {S_{2011}})}.
\label{fractional_var}
\end{equation}
 
\noindent ${S_{2010}}$ and ${S_{2011}}$ are the source fluxes measured in the deep 2010 and 2011 images respectively. $\hat{S}$ is the average of the 2010 and 2011 source fluxes. The errors on the 2010 and 2011 source fluxes are those measured in the deep images and calculated by {\sc aegean}.  The error on the fractional variability $\sigma_{f}$ is defined as the propagation of the flux errors through Equation \ref{fractional_var}. 

%%%%%%%%%%%%%%%%%%%%%%%%%%
\subsection{Transient searches}
%%%%%%%%%%%%%%%%%%%%%%%%%%
In conjunction with the variability searches described above, we also search the light-curves for short timescale (minutes to days) transient behaviour. The monitoring algorithms described in Section \ref{monitoring} give complete light-curves for all sources detected using forced measurements (estimates of the flux density at the position of a source in the absence of a blind detection). By including forced measurements into the light-curves, the variability metrics are capable of detecting transient sources. 

The variability search metrics are, however, less efficient at catching large changes in flux density over a small number of epochs. This is because the $\chi^{2}$ statistic is less sensitive to a small number of data points within a light-curve, that deviate significantly away from the weighted mean flux density of all the data points. We therefore search for this type of faster transient behaviour by examining all sources with a mean signal-to-noise ratio greater than 6.5$\sigma$, but that only have one to five detections. For a persistent source with a mean signal-to-noise ratio greater than 6.5$\sigma$, we would predominantly expect detections in all epochs.        
  
The VAST pipeline was also used to search for sources that have no known radio counterparts in any one of the following catalogues: VLSS, NVSS, PMN and Culgoora. This query was designed to search for long term transient activity whereby a source was not detected in the respective survey(s), but was subsequently detected in the MWA images. We did not search for sources that were not detected in the MWA images but were in the catalogues. 

%%%%%%%%%%%%%%%%%%%%%%%%%%
\section{Results}
\label{results}
\subsection{Science quality verification}
\label{science_quality _verification}
\begin{figure}
\centering
\includegraphics[scale=0.75]{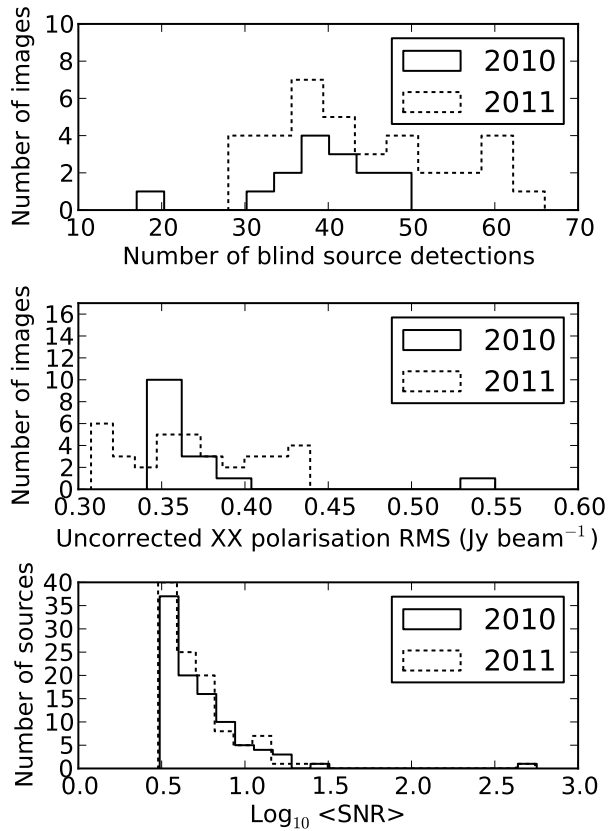}
\caption{Top panel: Histogram of the number of blind source detections through both expeditions. Note, sources that were not detected in previous images are automatically measured in later images (and vice versa), if not detected by the source finder. Middle panel: A histogram of the non-primary beam corrected XX polarisation RMS values of all the images from both 2010 and 2011. A mean RMS (calculated over all the images) of 0.36~Jy~beam$^{-1}$ is achieved before primary beam correction. Bottom panel: A histogram of the (log) mean signal-to-noise ratios of the source detections from 2010 and 2011.}
\label{source_counts}
\end{figure}

%%%%%%%%%%%%%%%%%%%%%%%%%%
At a source extraction threshold of 5.5$\sigma$, we find 105 sources within a solid angle of $\Omega = 1430$ deg$^{2}$. 
Figure~\ref{source_counts} (top panel) shows a histogram of the source detections per Stokes I image for both the 2010 and 2011 observations. From a total of 51 images, typically $\sim$40 blindly detected sources were found per image. The total number of extracted sources was a function of primary beam response and image RMS. The primary beam response changed between observations, so that differing populations of (weak) sources were found in different images. 

Note that the VAST monitoring algorithms only require a source to have a peak flux within an island of 5.5$\sigma$ in \emph{one} image (i.e. a single blind detection) to then be monitored in all images. Therefore the mean signal-to-noise ratio of source detections through all the images was in some cases lower than 5.5$\sigma$. The bottom panel of Figure~\ref{source_counts} shows a histogram of the (log) mean signal-to-noise ratios of source detections in the Stokes I images from 2010 and 2011 respectively. We remove from the analysis any source that had a mean signal-to-noise ratio lower than three. 

The middle panel of Figure \ref{source_counts} shows a histogram of the RMS values for the non primary beam corrected XX polarisation images. The RMS value for each image were calculated using an interquartile range that masked out bright sources. These pre-primary beam corrected RMS values from the calibrated images are shown because the Stokes~I images have non-uniform noise, after the primary beam has been divided out. 

To calculate the area of sky we have surveyed to a given sensitivity in the Stokes~I images, we produced sensitivity maps. This was achieved by multiplying the RMS values for each of the XX and YY polarisation images by the inverse of the XX and YY polarisation primary beam maps. These XX and YY polarisation sensitivity maps were then averaged together to produce a Stokes~I sensitivity map. Each pixel in the resulting Stokes~I maps therefore represents the 1$\sigma$ flux limit which the instrument was sensitive, at that location and time. 

\begin{figure}
\centering
\includegraphics[scale=0.58]{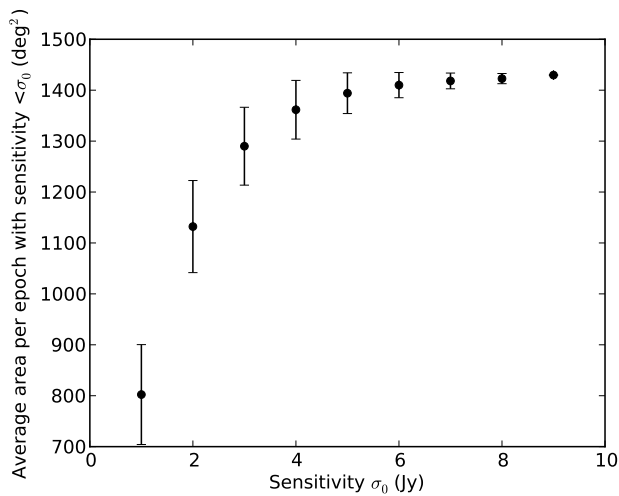}
\caption{A plot of average area of sky per epoch (y-axis) with sensitivity $\sigma_{0}$ or better (x-axis). The error bars represent one standard deviation of the areas over all the images.}
\label{area_vs_flux}
\end{figure}

For each Stokes~I sensitivity map, the sum of the number of pixels was calculated within a number of discrete sensitivity bins (see Table \ref{area_table} for bin widths). Each pixel corresponds to an area of sky 2.5$^{\prime} \times 2.5^{\prime}$ and the total amount of sky per image is $\Omega=1430$ deg$^{2}$. In Figure~\ref{area_vs_flux} we plot the average area per epoch to which we are sensitive to, with the sensitivity bins. In Table~\ref{area_table} we summarise the values shown in Figure~\ref{area_vs_flux}. At the most sensitive region of the primary beam, we observed a mean area of sky 802 $\pm$ 98 deg$^{2}$ to a 1$\sigma$ sensitivity $\leq$~1~Jy.
 
\begin{table}
\centering
\caption{Table of average area of sky sampled per epoch with sensitivity $\sigma_{0}$ or better. Also see Figure \ref{area_vs_flux}.}
\begin{tabular}{c|c|}
\hline  Sensitivity $\sigma_{0}$ (Jy ) & Area (deg$^{2}$) \\
\hline
1 &  802  $\pm$ 98 \\
2 & 1132 $\pm$ 90 \\
3 & 1290 $\pm$ 76 \\
4 & 1361  $\pm$ 58 \\
5 & 1394  $\pm$ 39 \\
6 & 1409  $\pm$ 24 \\
7  & 1418  $\pm$ 15 \\
8  & 1422  $\pm$ 10 \\
All &  1430  \\
\hline
\end{tabular}
\label{area_table}
\end{table}

\begin{figure*}
\centering
\includegraphics[scale=0.60]{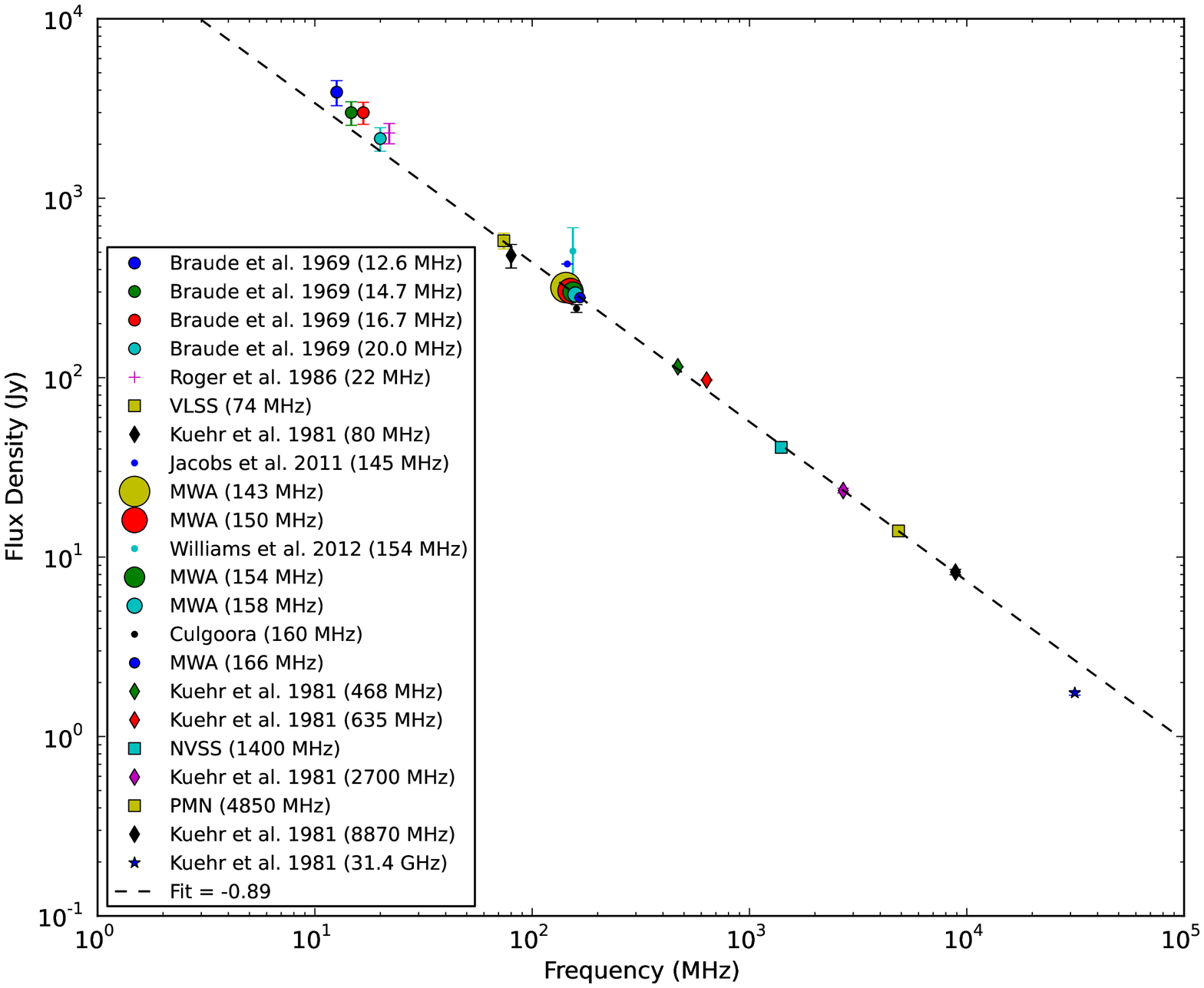}
\caption{Spectral energy distribution of Hydra A. The MWA fluxes (circles, various sizes) show the median Hydra A flux density values for each subband respectively. The 154 MHz MWA data point was calculated from the full continuum bandwidth image. Additional data points over a frequency range 10 MHz $-$ 31.4 GHz are shown (see \citealt{Braude1969}; \citealt{Danny2011}; \citealt{Kuehr1981}; \citealt{Roger1986} and \citealt{WIlliams2012} for full details). The data point of \citealt{WIlliams2012} is an independent reduction of the same data as considered here. The derived linear best fit using the VLSS, NVSS, PMN data points (which yields $\alpha=-0.89$ where $S_{\nu} \propto\nu^{+\alpha}$) is shown with a dashed line. This was the spectral index and derived flux scale used in the calibration procedure discussed in Section \ref{cal} .}
\label{HydA_SED}
\end{figure*}

Figure \ref{HydA_SED} shows the median fluxes of Hydra A, calculated from all four sub-band images (at 143, 150, 158 and 166 MHz) and from the full bandwidth image at 154 MHz, plotted as a function of frequency. In Figure \ref{HydA_SED} we also show other measurements of Hydra A from different instruments and surveys to evaluate the SED and relative flux scale. The closest measurement in frequency to the MWA-32T (154 MHz) observations are those from the Culgoora Circular Array at 160 MHz (see \citealt{Culgoora_Telescope}). Note that the Culgoora Circular Array has a maximum baseline length of $B_{max}\sim3.39$ km \citep{Culgoora_Telescope} whereas the MWA-32T extent is $B_{max}\sim350$m. Hydra A is slightly resolved even at MWA-32T baseline lengths, therefore we expected to get a higher flux density than that reported by \cite{Culgoora_CAT}, because the Culgoora Circular Array was more sensitive to the resolved diffuse structure only (see \citealt{Lane2004}). We derived a spectral index $\alpha = -0.89$ (where $S_{\nu} \propto\nu^{+\alpha}$) from a best fit to the VLSS, NVSS and PMN survey data points (shown with a dashed line). This spectral index and corresponding flux scale were used for the calibration of the MWA-32T data. The flux scale is in agreement with historical measurements of Hydra~A at a variety of frequencies.   

\begin{figure}
\centering
\includegraphics[scale=0.70]{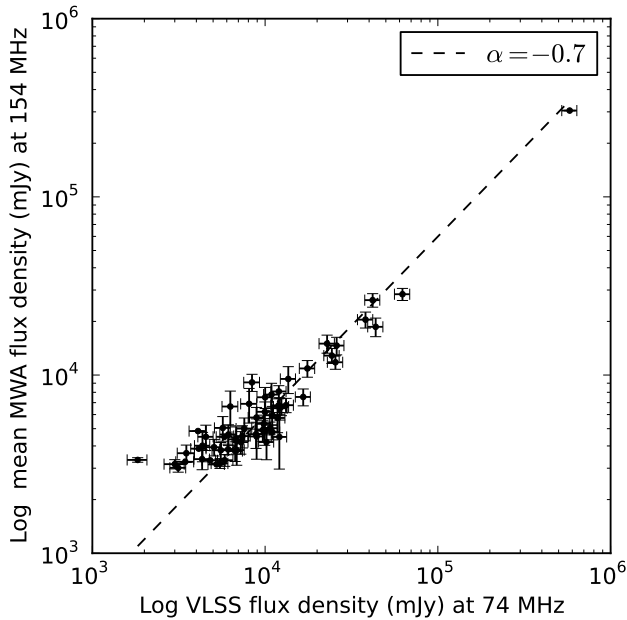}
\caption{Mean MWA source flux density at 154 MHz plotted against cross-matched VLSS flux density at 74 MHz. The dashed line shows the trend that would be expected if all the sources had a spectral index of $\alpha=-0.7$.}
\label{VLSS_flux_comp}
\end{figure}

In Figure \ref{VLSS_flux_comp} we compare the fluxes of the extracted sources from observations in 2010 with the crossmatched fluxes of sources in the VLSS catalogue \citep{VLSS}. The correlation that would be expected if the sources all had spectral indices $\alpha=-0.7$ is shown by a dashed line. We also repeated this analysis using the NVSS and Culgoora catalogues. 
%In Figure~\ref{Col_flux_comp} we plot the same relationship for the Culgoora catalogue \citep{Culgoora_CAT}. In this case the survey frequencies are very close (154 MHz MWA and 160 MHz Culgoora) but the resolutions of the instruments were different. 
We found that the MWA fluxes were in good agreement with the crossmatched catalogue fluxes (noting the differences in frequency and instrument type). For further discussion on the flux densities of Southern Hemisphere low frequency radio sources see \cite{Danny2011}.

\begin{figure}
\centering
\includegraphics[scale=0.52]{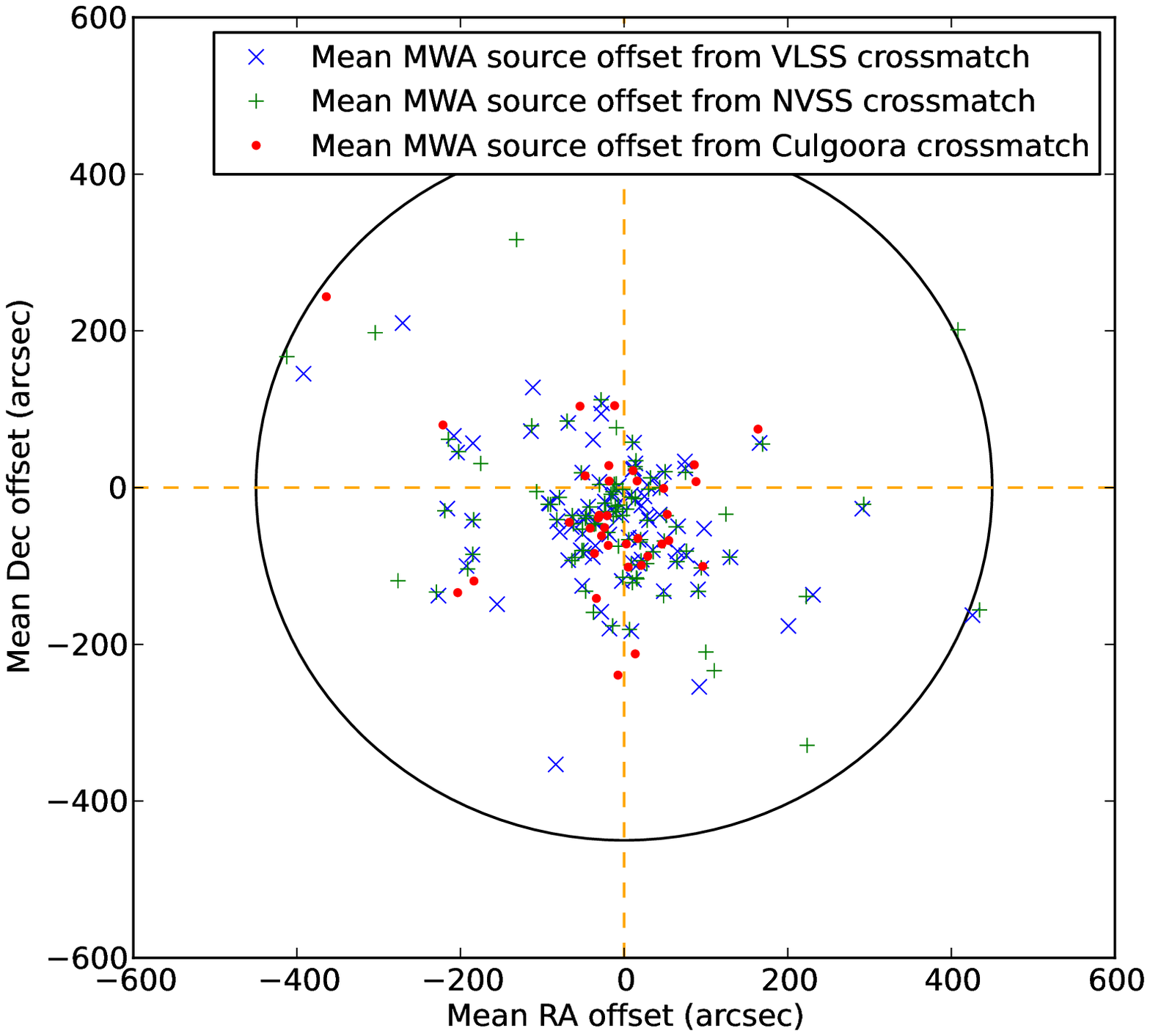}
\begin{tabular}{c|c}
\includegraphics[scale=0.36]{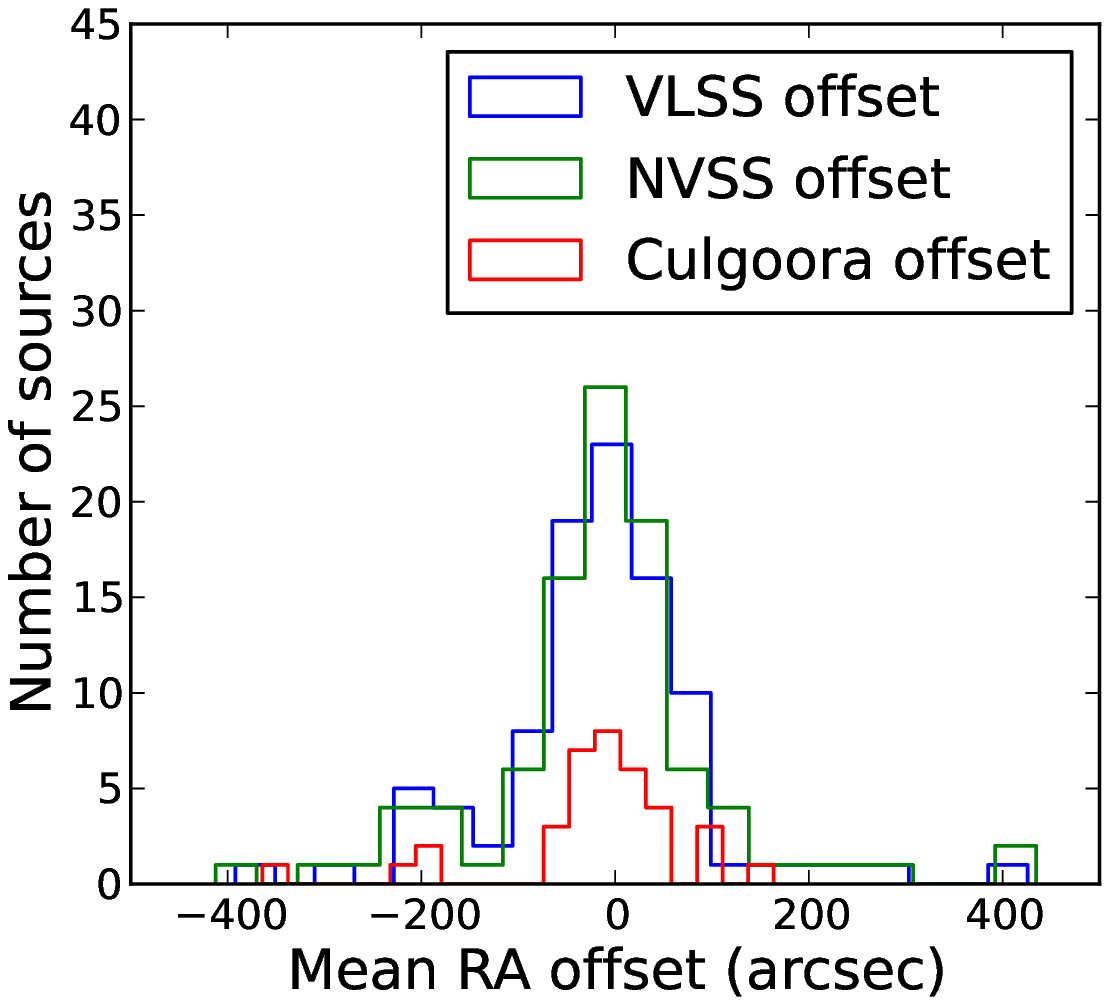} 
\hspace{-0.2in}
\includegraphics[scale=0.36]{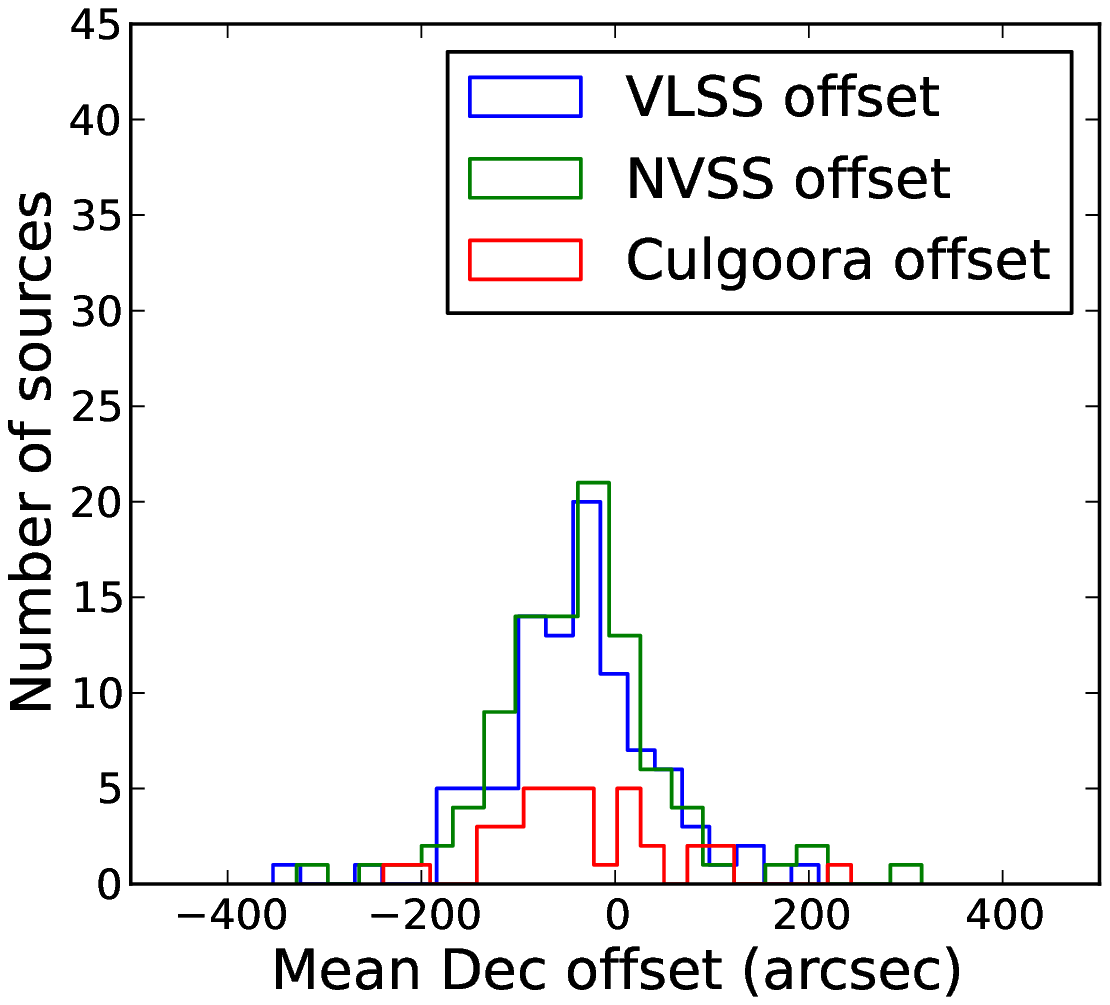}  
\end{tabular}

\caption{The top panel shows a plot of the average RA and Dec offset per-source from the cross-matched catalogue counterpart (either VLSS, NVSS or Culgoora). The RAs and Decs are averaged per-source over the total number of images (in 2010 only). The differences are then measured between the average MWA positions and the corresponding catalogue positions. The black circle denotes a typical synthesised beam diameter of 15$^{\prime}$. The bottom panels show histograms of the RA and Dec offsets (per survey).}
\label{Positional_offsets}
\end{figure}

In addition to examining the fluxes of the extracted sources with respect to other survey catalogues, we also examined the positional accuracy. Figure \ref{Positional_offsets} (bottom panel) shows the average MWA source position through all of the images, subtracted from the crossmatched catalogue position, yielding an average positional MWA offset.  In Figure~\ref{Positional_offsets} we also plot a typical beam diameter of 15$^{\prime}$ (black circle), showing that the mean positions of all the sources all lie within one beam diameter of the catalogue sources (NVSS, VLSS and Culgoora). On the top panel of Figure \ref{Positional_offsets} we show histograms of MWA positional offsets. There is a slight offset in declination ($\sim20^{\prime \prime}$) from the catalogues. This offset could potentially be due to a instrumental pointing offset, or differences in the defined position of Hydra~A (used for calibration) with respect to the catalog position.   

\begin{figure}
\centering
\includegraphics[scale=0.65]{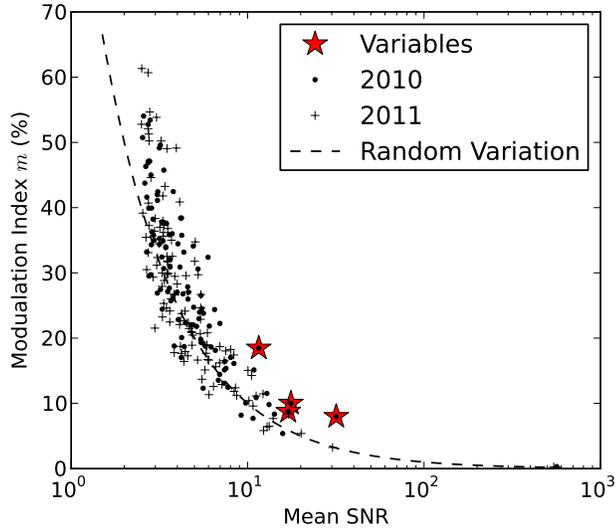}
\caption{The modulation indices for all sources in 2010 (dots) and 2011 (crosses) as a function of $\overline{SNR}$. The dashed line shows the variability that would be expected from random variations as derived from the measurement uncertainty. The red stars denote the variable sources that are discussed in Section \ref{discussion}.}
\label{V1}
\end{figure}

In Figure \ref{V1} we plot the modulation indices for all sources from 2010 (dots) and 2011 (crosses) as a function of the log mean signal-to-noise ratio ($\overline{SNR}$). We also plot the random variability that would be expected as a function of $\overline{SNR}$ (dashed line). The modulation indices of the sources within the 2010 and 2011 observations adhere broadly to the correlation that would be expected due to the random variation of the sources. This is a diagnostic plot to check that the actual variations of the sources are in agreement with the predicted variations (based on the mean signal-to-noise ratio of the source detection).     

From the results above, we conclude that the flux scale of Hydra~A is set accurately with respect to observations reported in the literature. The flux scale of the ensemble of sources agrees well with catalogue cross-matches from a number of surveys, and the cross-matches are positionally accurate.  

%%%%%%%%%%%%%%%%%%%%%%%%%%%%%
 \subsection{Variability search results on timescales of minutes to days}
\begin{figure}
\begin{tabular}{c}
 \includegraphics[scale=0.62]{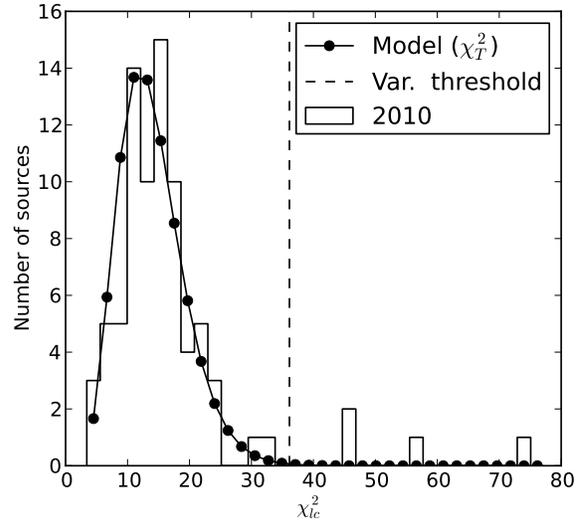}  \\
 \includegraphics[scale=0.62]{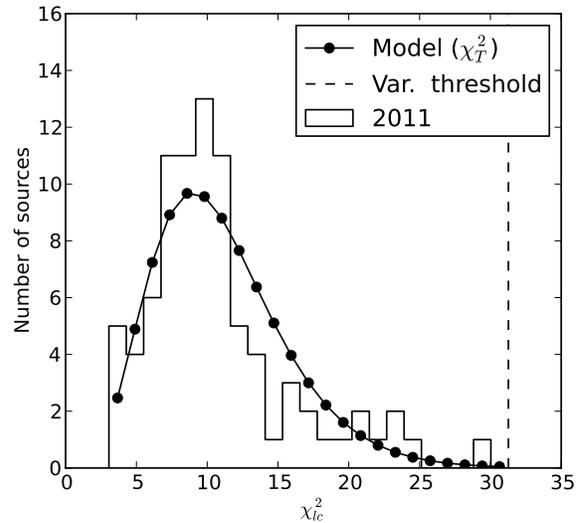}  
\end{tabular}
\caption{Top panel: A histogram of the $\chi^{2}_{lc}$ values from the 2010 light-curves. The solid line shows the theoretical distribution that would be expected with 14 degrees of freedom, and is calculated at the same bin centres as the histogram. The vertical dashed line shows the variability detection threshold of $\chi^{2}_{lc}=36.1$, all sources to the right of this line  are deemed variable. Bottom Panel: The same plot but for the 2011 data. The 2011 analysis uses a reduced set of data points with 11 degrees of freedom. The vertical dashed line shows the variability detection threshold of $\chi^{2}_{lc}=31.2$.}
\label{chis}
\end{figure}

 %%%%%%%%%%%%%%%%%%%%%%%%%%%%%
 \label{var_search}
The source light-curves from 2010 and 2011 were searched for variability using the methods discussed in Section \ref{var_section}.  
The top panel of Figure \ref{chis} shows a histogram of the $\chi^{2}_{lc}$ values of the light-curves in 2010. The solid dotted line shows the theoretical distribution $\chi^{2}_{T}$ of  values that would be predicted based on the number of degrees of freedom, and the number of light-curves. The vertical dashed line shows the variability detection threshold of $\chi^{2}_{lc} > 36.1$ (or $P<0.001$). Sources that are above this threshold are considered variable. Four sources are identified as variable and we will discuss these later in this section.       

When initially examining the histogram of the $\chi^{2}_{T}$ distribution of sources for the 2011 data, it appeared skewed to the left of the theoretical distribution. Extensive testing of the 2011 dataset revealed that observations taken closer together in time ($<$10 mins) had a high degree of correlation between image pairs (after bright sources had been masked out). The $\chi^{2}$ analysis relies on the noise properties being Gaussian and un-correlated between individual images. 

The observing strategies differed between the 2010 and 2011 observations. The 2010 observations were spaced (fairly consistently) $\Delta T \sim$30 minutes apart (see Figure \ref{hist}), with hour angles between $-1.68$ and 2.13 hours. In 2011, the observations were predominantly spaced $\Delta T <$10 minutes apart, with hour angles between $-1.61$ and 3.06 hours. Similar beam-former settings (which determines the primary beam response) were also typically used for all the observations in 2011, whereas in 2010 they were different from each other. We attribute the correlation observed in the 2011 data to increased side-lobe confusion (including source confusion) within the image pairs. The correlation increased for observations that shared the same beam-former settings, were spaced close together in time, and were observed at large hour angles.  

The initial skewing of the distribution, with more lower values than expected, confirmed that the correlated images were affecting our ability to use the $\chi^{2}$ statistic. To account for this we removed the highly correlated 2011 image pairs (with  $\Delta T < 10$ minutes) from the analysis. This reduced the number of images in the 2011 sample from 36 down to 12. The bottom panel of Figure \ref{chis} shows the resulting $\chi^{2}_{lc}$ distribution from the reduced 2011 sample (after the correlated images were removed). The distribution fits reasonably well with the theoretical distribution and the analysis identifies no variable sources above the variability detection threshold of $\chi^{2}_{lc} > 31.2$ (or $P<0.001$).   

Four sources meet the criteria of being variable using the analysis presented above. We note that all these sources were deemed variable based solely on the 2010 observations; no sources were found to be variable in the 2011 observations. The sample of observations from 2011 used for the variability analysis had less observations than 2010, and the flux density measurements from those images often had larger uncertainties. This may have degraded our ability to to detect variability in the 2011 observations. 

\begin{figure*}
\vspace{-0.2in}
\includegraphics[width=4.5in, angle=90]{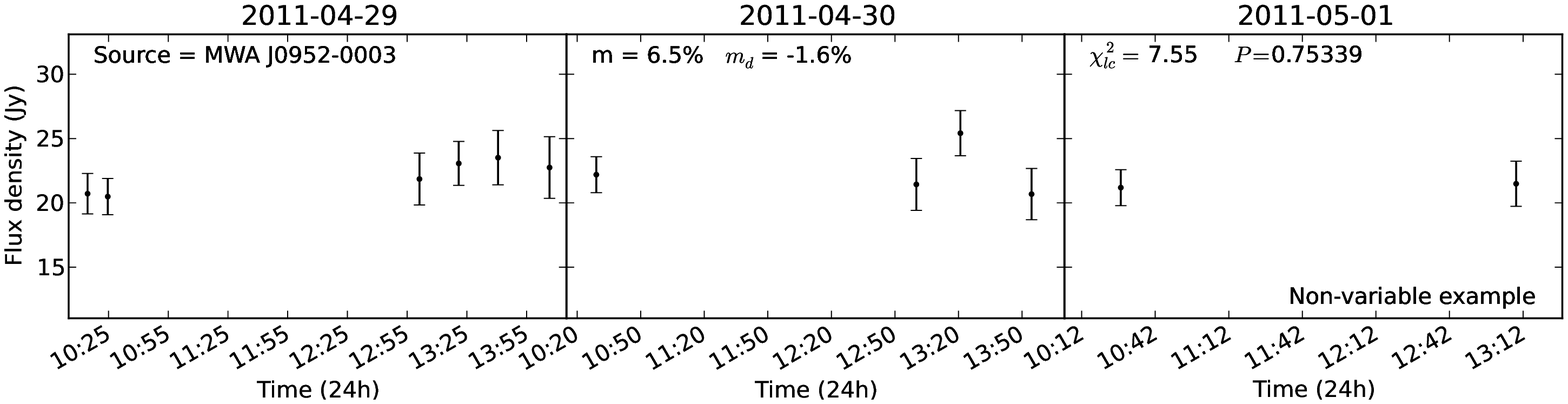} 
\includegraphics[width=4.5in, angle=90]{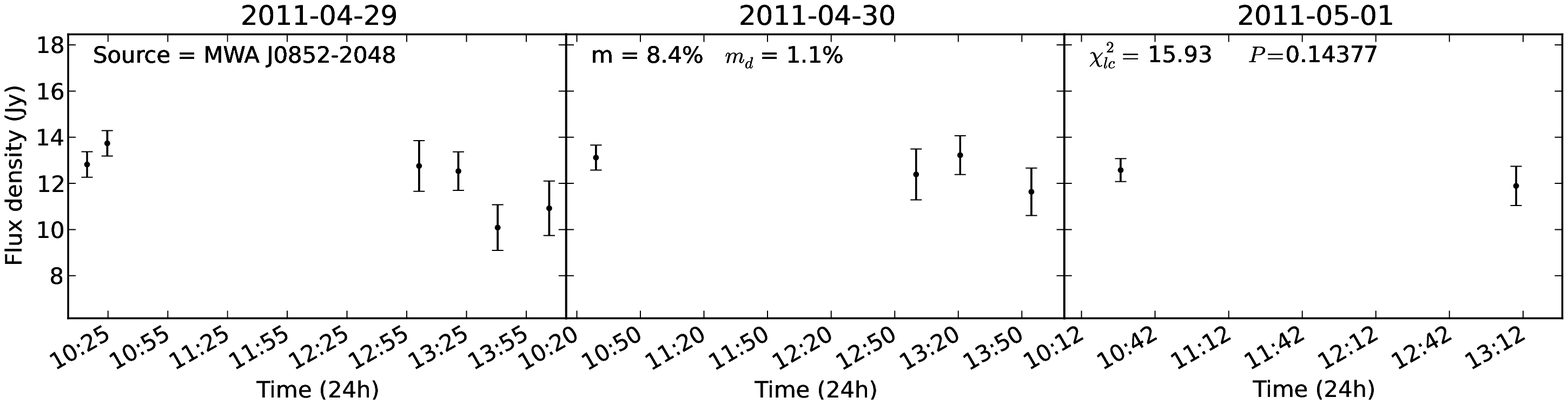}
\includegraphics[width=4.5in, angle=90]{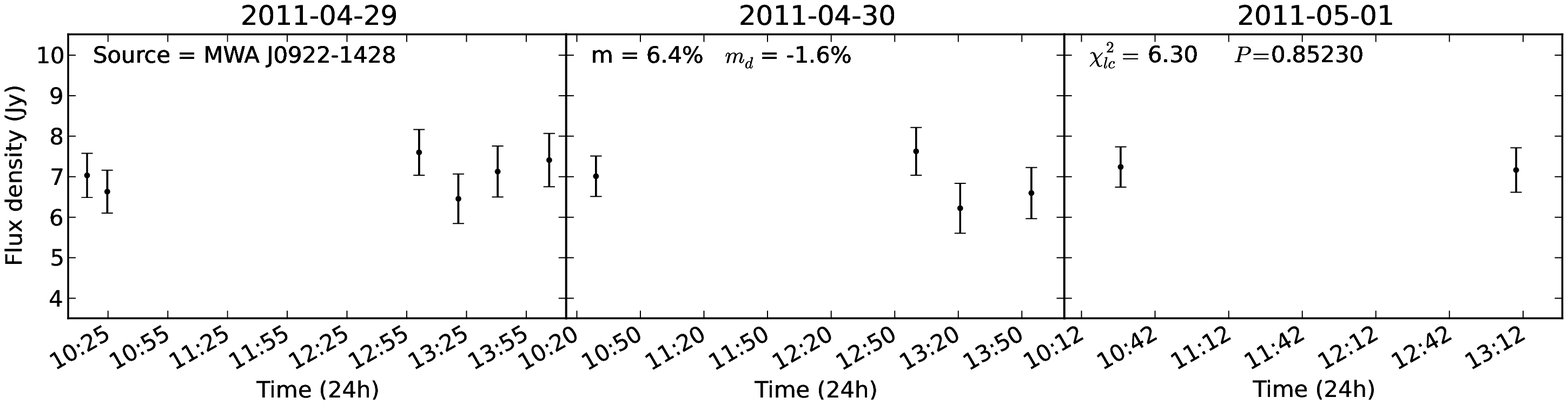} 
\includegraphics[width=4.5in, angle=90]{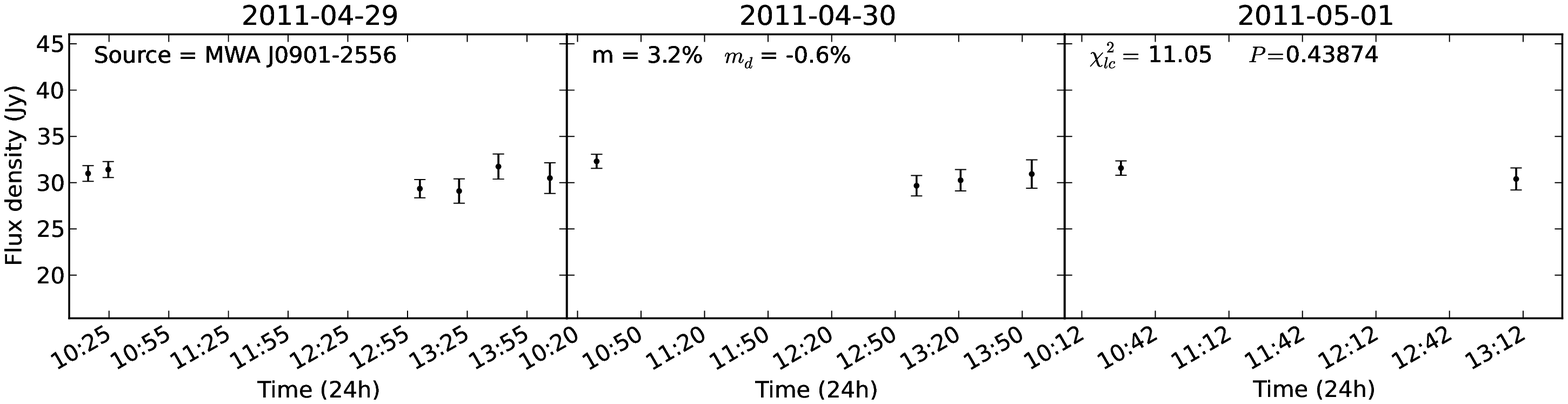} 
\includegraphics[width=4.5in, angle=90]{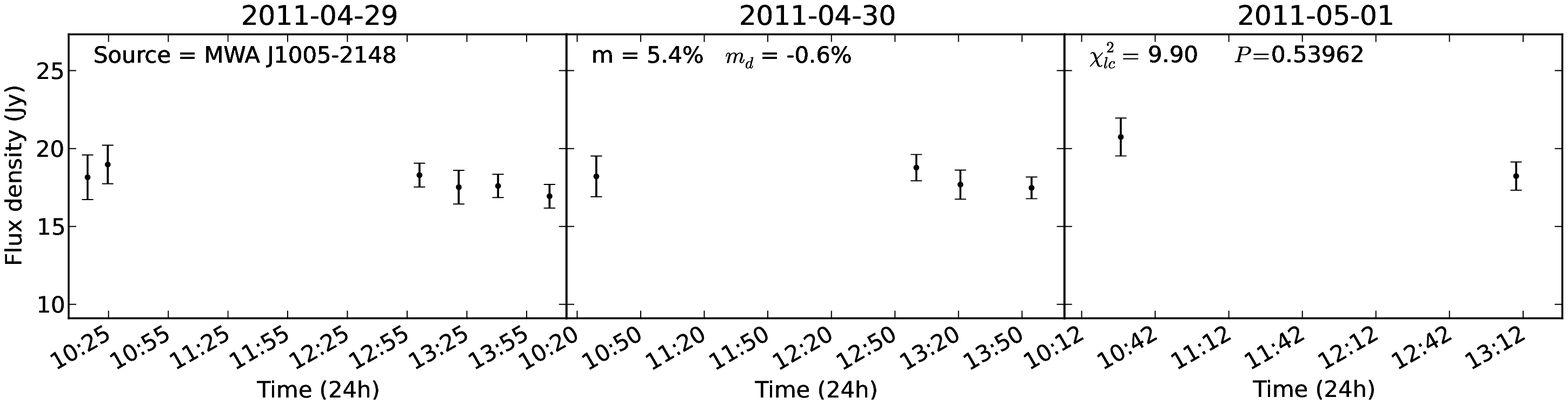} \\

\includegraphics[width=4.5in, angle=90]{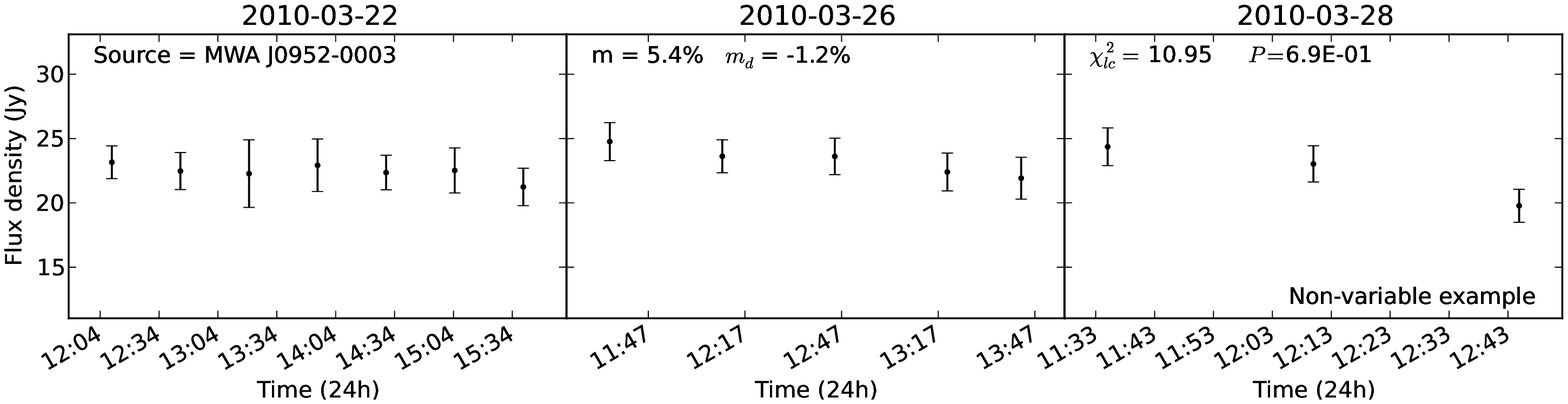} 
\includegraphics[width=4.5in, angle=90]{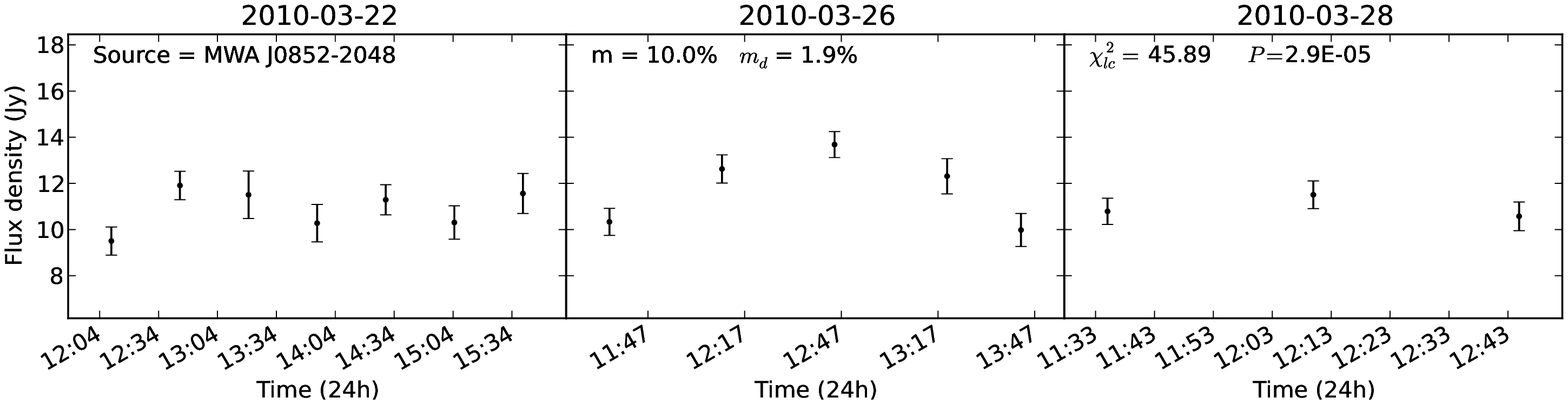}
\includegraphics[width=4.5in, angle=90]{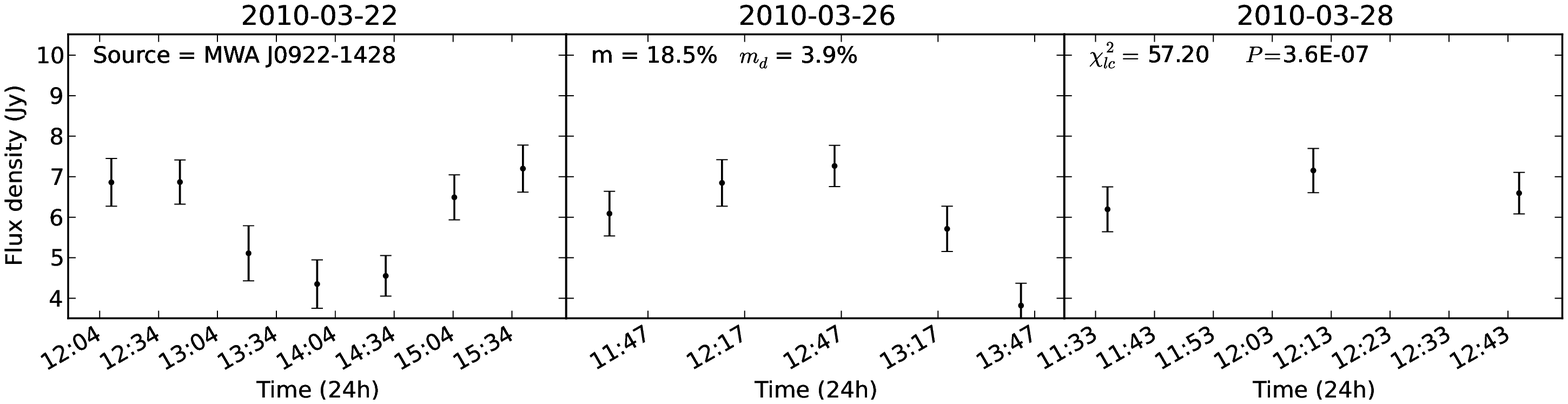} 
\includegraphics[width=4.5in, angle=90]{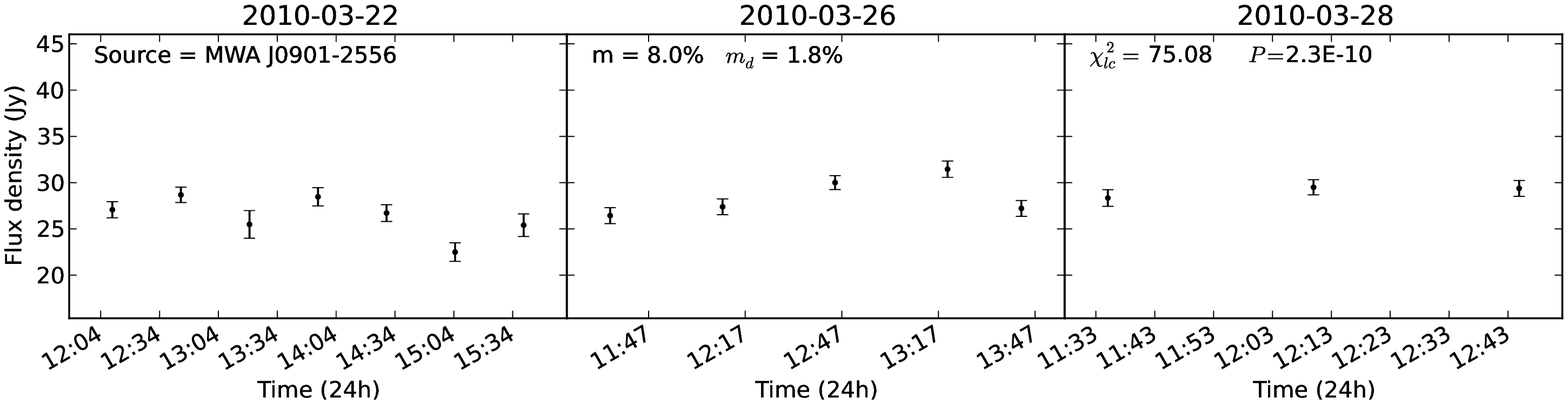} 
\includegraphics[width=4.5in, angle=90]{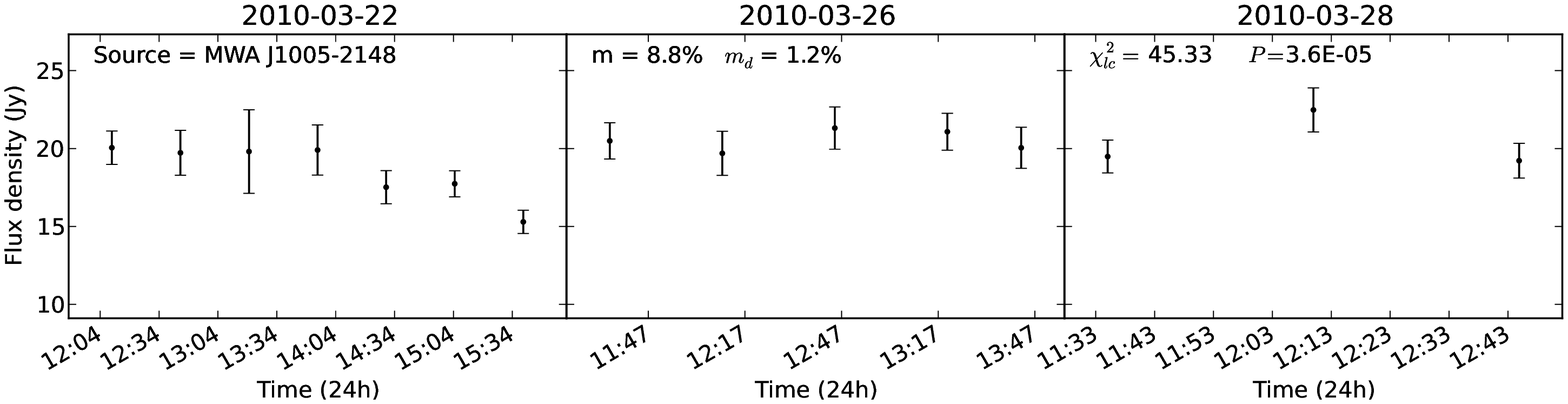} 
%\vspace{-0.2in}
\caption{Daily light-curves for sources identified as variable. Note that the top row shows an example non-variable source (MWA J0952-0003).
The figures on the left column show the daily light-curves for observations from 2010. The corresponding source light-curves for 2011 are show on the right column. The source statistics are indicated for each row and panel.}
\label{Example_lightcurves}
\end{figure*}

The statistics for these sources are summarised in Table~\ref{var_stats}. The light-curves for these sources are shown in Figure~\ref{Example_lightcurves}. Also included in Figure \ref{Example_lightcurves} in an example non-variable source (MWA J0952-0003).  The variability within these sources on the whole is relatively low, with  $1.2 \% < m_{d} < 3.9 \% $ (2010 observations) and $-0.6 \% < m_{d} < 1.1 \% $ (2011 observations). For each of the variable sources we will disucss the multi-wavelength properties below. \\

\noindent {\bf MWA J0852-2048}: Identified as radio galaxy PKS~B0850-20. There is no known X-ray, $\gamma$-ray counterpart. The radio galaxy is at a redshift $z=1.337$ \citep{Phil_Best}. The spectral index derived from a best fit to the VLSS, MWA and NVSS data points is $\alpha =-0.77$. \\

\noindent {\bf MWA J0922-1428}: Identified as the quasar QSO~B0919-142 with spectral index between 408 MHz and 1410 MHz of $\alpha =-0.83$ \citep{Quiniento}. The Quasar is at a redshift $z=0.305$ \citep{1993MNRAS.263..999T}. The spectral index derived from a best fit to the VLSS, MWA and NVSS data points is $\alpha =-0.6$. There is no known X-ray or $\gamma$-ray counterpart.  \\

\noindent {\bf MWA J0901-2256}: Identified as quasar LEDA~2825724 \citep{LEDA}. 
It also has an X-ray counterpart (1RXS J090129.1-255357) and near infrared galaxy (DENIS-P J090152.0-255333; \citealt{Dennis}) within the positional errors. The spectral index derived from a best fit to the VLSS, MWA and NVSS data points is $\alpha =-0.72$. \\

\noindent {\bf MWA J1005-2148}: Identified as radio galaxy LEDA 2826147  \citep{LEDA}. This source is also consistent with the X-ray source 1RXS J100511.0-214500. The spectral index derived from a best fit to the VLSS, MWA and NVSS data points is $\alpha =-1.2$.

\begin{table*}
\centering
\caption{Summary of modulation and de-biased modulation indexes of the variable sources detected in the daily images.}
\begin{tabular}{r|r|r|r|r|r|r|r}
\hline  \multicolumn{1}{c}{Name} &  $m$ (\%) & $m$ (\%)  &  $m_{d}$ (\%) & $m_{d}$ (\%) & \multicolumn{1}{c}{$\chi^{2}_{lc}$} & \multicolumn{1}{c}{$\chi^{2}_{lc}$}  \\
 &   (2010) &  (2011) & (2010) & (2011) & (2010) & (2011)  \\
\hline
MWA J0852-2048 &    10.0 & 8.4  & 1.9  & 1.1 & 45.89 & 15.93 \\
MWA J0922-1428 &    18.5 & 6.4  & 3.9  & -1.6 & 57.20 & 6.30 \\
MWA J0901-2556 &    8.0 & 3.2  &  1.8  & -0.6 & 75.08 & 11.05\\
MWA J1005-2148 &    8.8 & 5.4  &  1.2  & -0.6 & 45.33 & 9.90\\
\hline
\end{tabular}
\label{var_stats}
\end{table*}

%%%%%%%%%%%%%%%%%%%%%%%%%%
\subsection{Variability search results on the timescale of one year}
\begin{figure}
\centering
\includegraphics[scale=0.60]{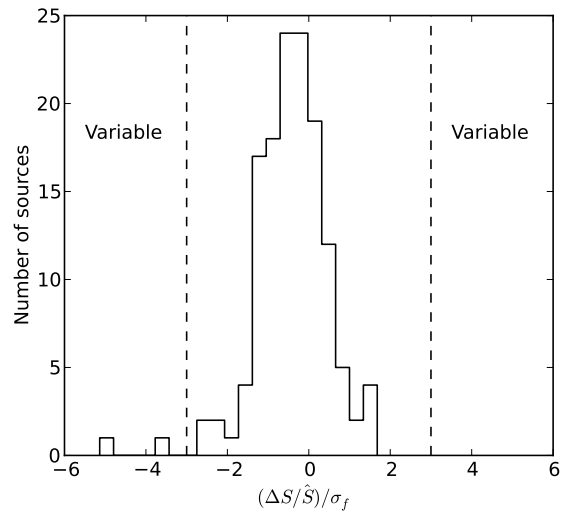}
\caption{A histogram of the fractional variability measured between the deep 2010 and 2011 images, normalised by the combined uncertainly from both years ($\sigma_{f}$). A source is deemed is variable if $|\Delta S/ {\hat S}|/ \sigma_{f} > 3$ and these regions are indicated in the figure.}
\label{hist_year}
\end{figure}

%%%%%%%%%%%%%%%%%%%%%%%%%%
The deep images created from the averages of the 2010 and 2011 data (respectively) were searched for variability, to probe timescales of one year. From the deep images, 136 sources were detected with signal-to-noise ratios greater than 5.5$\sigma$. In Figure \ref{hist_year} we show a histogram of the fractional variability $f = \Delta S / {\hat S}$, normalised by the combined uncertainly from the 2010 and 2011 deep images $\sigma_{f}$ (see Section \ref{var_section} for definition). We consider a source to be significantly variable if $|\Delta S / {\hat S}| > 3\sigma_{f}$. We find two sources that meet this criteria and we will discuss these in Section \ref{discussion}. 

We note that there is a slight skew in the data away from a mean of zero. The peak pixels of Hydra~A (flux calibrator) from the 2010 and 2011 images agree  to within 0.065\%. The peak fluxes from the Gaussian fits, however, only agree to 0.23\%. We assumed a beam size derived from a fit to Hydra A as the beam properties of the image. Slight differences in the beam properties and hence the derived fluxes may explain the offset in the histogram in Figure \ref{hist_year}. 

%Note that these images were dominated by confusion noise. Stacking all the images per-year can yield improved averaged flux measurements of the sources. This will not, however, improve the noise in the image nor the the error budget (derived from the image RMS noise) on the flux measurements. 

%%%%%%%%%%%%%%%%%%%%%%%%%%
\subsection{Transient search results}
%%%%%%%%%%%%%%%%%%%%%%%%%%
No short timescale transients were detected from the daily images. Furthermore, no longer duration transient sources were detected between the 2010 and 2011 deep images.  

%%%%%%%%%%%%%%%%%%%%%%%%%%
\section{Discussion}
%%%%%%%%%%%%%%%%%%%%%%%%%%
\label{discussion}
\subsection{Variables on timescales of minutes to days} 
\label{daily_var_discussion}
Four variable radio sources were found that showed significant variability. The mean time difference between consecutive observations was 26.0 minutes. All of these sources showed relatively low levels of variability. We find no sources (on these timescales) within our sample that showed significant variability with modulation indices greater than 50\%. There are 20 sources with mean fluxes greater than 6~Jy and of these, four are found to be variable. 

To assess if the variability is driven by the synchrotron process and intrinsic to the source, we can use the brightness temperature as a diagnostic tool (e.g. \citealt{BT_equation}). 
The brightness temperature of an unresolved source can be calculated using the following expression:

\begin{equation}
T_{B} = \dfrac{S\lambda^{2}}{2k_{B} c^{2}\Omega},
\label{Tb_eq}
\end{equation}

\noindent where $S$ is the flux density, $\lambda$ is the wavelength, $k_{B}$ is the Boltzmann constant and $\Omega$ is the solid angle subtended by the source. The solid angle can be approximated using $\Omega = r^{2}/D^{2}$, where $r$ is the radius of the source and $D$ is the distance. We assume that the variability timescale $\tau$, and hence the light travel time, is the limiting factor on the source size i.e. $r=c \times \tau$. This would therefore give a minimum brightness temperature of  

\begin{equation}
T_{B} \geq \dfrac{\Delta S\lambda^{2}D^{2}}{2k_{B} c^{2}\tau^{2}},
\label{Tb_eq}
\end{equation}

\noindent where $\Delta S$ is the amplitude of variability. Taking the source MWA~J0852-2048 as an example and substituting $\Delta S =$ 1~Jy, $D = 9.5$~Gpc (calculated from a redshift of $z=1.337$) and $\tau$~= ~26 minutes we find $T_b > 5.2\times10^{32} K $. The timescale of variability within this source is therefore not consistent with intrinsic variability, as it would break the Compton catastrophe limit \citep{Kell1969} of $T_b < 10^{12} K $. Doppler boosting from the relativistic beaming of jets can increase the brightness temperature of a given source. Following \cite{BT_paper} where the increase in brightness temperature due to Doppler boasting is defined as being proportional to $D^{3}$ (where D is the Doppler boosting factor), and assuming a generous value of $D=20$, we find the pre-Doppler boosted brightness temperature of MWA J0852-2048 to be $T_b > 6.5\times10^{28} K $. For incoherent sources, even with generous modification of the parameters used in Equation \ref{Tb_eq} and taking into account Doppler boosting, it is extremely difficult to bring the brightness temperature below the required Compton catastrophe limit.

Two of the four variable sources are associated with QSOs, which are a population of sources that typically have compact radio emitting regions and potentially could undergo refractive interstellar scintillation. The time scales for refractive scintillation are, however, typically much longer than those measured: on the order of months to years (\citealt{Hunstead_72}). Diffractive scintillation can also be ruled out because the angular sizes of the proposed crossmatched counterparts (radio galaxies, QSO) are too large. At these frequencies, compact objects the size of pulsars (with angular size $\sim 10^{-9}$ arc seconds; \citealt{Lazio2004}) typically undergo diffractive scintillation (see also \citealt{Rickett77}). Inter-planetary scintillation is also another possible explanation, however, the timescales are typically no longer than minutes (see \citealt{Cohen67} and \citealt{Cohen67b}). The solar elongations for these observations were typically $\theta >90^{\circ}$, where in comparison, inter-planetary scintillation is more prominent with $\theta < 20^{\circ}$ \citep{Cohen67}. 

The mechanism responsible for intra-day variability (IDV) at GHz frequencies cannot be invoked to explain IDV in quasars at  $\sim 150$ MHz.  It is now well established that the principal mechanism responsible for IDV in the frequency range 1-10 GHz is interstellar scintillation (e.g. Lovell et al. 2008 and references therein). For lines of sight off the Galactic plane the scattering occurs in the regime of strong scintillation at frequencies below $\sim$2-5\,GHz \citep{1998MNRAS.294..307W}, and the flux density variations are caused by refractive interstellar scintillation (RISS; \citealt{1995A&A...293..479R}; \citealt{1997ApJ...490L...9K}, \citealt{2000ApJ...529L..65D}; \citealt{2000ApJ...538..623M}).  The timescale associated with RISS scales as $\lambda^{\beta/(\beta-2)}$, where $\beta$ is close to the value for Kolmogorov turbulence of $11/3$ \citep{1995ApJ...443..209A}.  Thus the characteristic timescale of RISS for a source which exhibits intra-day variations at 2\,GHz is $\approx 300\,$days at 150\,MHz.  Such slow RISS is manifest in the phenomenon of low frequency variability first identified in AGN by \cite{Hunstead_72} and interpreted by \cite{1984A&A...134..390R}.

Ionospheric scintillation could potentially explain the variability seen within these sources. Ionospheric phase errors are typically more common and prominent than amplitude related errors \citep{Jacobson}. However, phase errors that manifested themselves in the image plane could have affected source fitting and introduced low levels of variability (between images). For example, smearing and deformation of point sources will result in the reduction of the measured peak fluxes, which in turn will introduce apparent variability between images (e.g. see \citealt{Intema2009}; \citealt{Cotton}). 

Sources separated by more than one degree in distance and one minute in time often require distinct calibration solutions \citep{Braun}. For example, \cite{Bernardi_2010} found three bright radio sources 1$-$2 degrees from the phase centre of their observations of 3C196 at 150 MHz, from the Westerbork Radio Synthesis (WSRT) telescope, that showed imaging artefacts caused by ionospheric effects. They reported that the ionospheric turbulence occurred on the timescale of minutes and that the degree of ionospheric scintillation varied from night to night.  Also, different lines of sight within the primary beam ($12^{\circ} \times 12^{\circ}$) experienced different regions of turbulence. 
 
For this survey we only generated a single time-independent calibration solution in the direction of Hydra~A. The simplification of the ionospheric complexity could be responsible for the variability reported. The connection between ionospheric variations and amplitude variations within the sources is still however rather tentative. The real-time system \citep{Mitch_2008} will provide direction dependent calibration and ionospheric modelling in MWA future operations, which will minimise ionospheric effects.   

The variability seen could be ascribed to some un-diagnosed instrumental or calibration effect, however, a number of sources that have similar fluxes and positions in the primary beam remained non-variable over the same timescales. For example, the variable source MWA J0901-2556 is the second brightest in the field ($\overline{S} = 30$~Jy) and sits at the edge of the image ($r=14.1$ degrees from the pointing centre). In comparison the source MWA J0952-0003 (see Figure \ref{Example_lightcurves}) which is the third brightest in the field ($\overline{S} = 22$~Jy), is non-variable and also sits at the edge of the image ($r=14.5$ degrees from the pointing centre).  

The effect would therefore have to be direction dependent and fairly localised. As discussed in this paper, approximations were made in assuming Hydra~A was a point source, and also in using a monochromatic primary beam solution over the entire bandwidth. These, plus potentially other sources of error, cannot be ruled out as causes for the low levels of variability reported in these four sources. We conclude that it is unlikely that the variability is caused from an astrophysical process. 
In Section \ref{density} we compare and discuss the upper limit on the surface density of these four variable sources, with other surveys reported in the literature. 

%%%%%%%%%%%%%%%%%%%%%%%%%%%%%%%
\subsection{Variables on timescales of one year}
%%%%%%%%%%%%%%%%%%%%%%%%%%%%%%%
By comparing the fluxes of sources detected in the 2010 and 2011 deep images, we have probed variability on timescales of $\sim$404 days. Within our sample we find two sources that show significant variability (with $|\Delta S/{\hat S}| >$ 3$\sigma_{f}$). The source MWA~J0837-1950 had a fractional variability $f=28.6$\%. This source is associated with an AGN (LEDA 2825560; \citealt{LEDA_89}) and the radio spectral index calculated from the VLSS, MWA and NVSS is fairly flat $\alpha =-0.1$. The interpretation of the variability of this source is consistent with long term intrinsic AGN variability. It is also plausible that refractive scintillation has played a role in the flux density change. 

The source MWA~J0901-2556 had a fractional variability $f=10.0$\% on the timescale of one year. This source was also identified as being variable from the analysis of the daily images (see section \ref{daily_var_discussion}). For this source we previously ruled out a number of variability mechanisms for short timescales. It is possible that two different mechanisms are independently causing the variability on the timescales of 26 minutes and one year respectively. Intrinsic variability and refractive scintillation are plausible explanations for the long term variability. As discussed previously ionospheric effects or calibration errors could be causing the short timescale variability.    

In comparison to previous work, \cite{Mcgilchist_1990} reported that 1.1\% of their sample of 811 sources with $S>0.3$ Jy show a magnitude of fractional variability $\Delta S/{\hat S}$ $>15\%$, over timescales of one year at 151 MHz. The significance of the variable sources reported by \cite{Mcgilchist_1990} was  $|\Delta S/{\hat S}| >$ 3$\sigma_{f}$. \cite{Mcgilchist_1990} find no sources with $S>0.3$~Jy that show fractional variability $>4\%$. 

\cite{slee_1988} report on the variability of 412 radio sources with $S>3.0$~Jy on timescales of months and years at 160 MHz. Using monthly data, \cite{Mcgilchist_1990} calculate that 13.3\% of the \cite{slee_1988} sample show fractional variability at a level  $>15\%$. Two of our 136 sources show significant variability on the timescale of one year, which equates to 1.5\% of our sample. The results from our study are therefore in better agreement with those of \cite{Mcgilchist_1990} than those of \cite{slee_1988}. 

It should be noted however that there are differences between the samples used to calculate the statistics described above. In the \cite{Mcgilchist_1990} sample there were only $\sim$20 sources with $S>3$~Jy, whereas in our sample we have 136 sources and in the \cite{slee_1988} sample there were 412. Furthermore, the typical flux error in the \cite{Mcgilchist_1990} sample was $\sim$10 mJy, whereas in our sample it was $\sim$300 mJy. The higher RMS in our sample obviously degrades our ability to detect significant variability. There were also subtle differences in the instruments and strategies that were used to take the data e.g. resolution and cadence.      

 %%%%%%%%%%%%%%%%%%%%%%%%%%
\subsection{Surface density of transients and variables}
%%%%%%%%%%%%%%%%%%%%%%%%%%
\label{density}
No transients were detected on short or long timescales. We can therefore place upper limits on the surface density (sometimes referred to as 2-epoch snapshot rate or areal density) of low frequency transient sources over the whole sky. A source extraction level of $5.5\sigma$ was used and Table~\ref{rho_table} summarises the average area of sky over which transient sources could have been detected (see also Section \ref{science_quality _verification} for details). 

\begin{table}
\centering
\caption{Table summarising the mean area of sky of which we were capable of detecting transients, calculated at five flux levels.
The last column shows the surface density upper limits ($\rho$) calculated from the sky areas respectively.}
\begin{tabular}{c|c|c|c|}
\hline  Transient flux (Jy) & Area (deg$^{2}$) & $\rho$ (deg$^{-2}$) \\
\hline
~5.5 &  802  $\pm$ 98 & 7.5 $\times 10^{-5}$\\
~~11 & 1132 $\pm$ 90 & 5.3 $\times 10^{-5}$  \\ 
16.5 & 1290 $\pm$ 76  & 4.6 $\times 10^{-5}$   \\
~~22 & 1361  $\pm$ 58 & 4.4 $\times 10^{-5}$ \\
49.5 &  1430  & 4.2 $\times 10^{-5}$\\
\hline
\end{tabular}
\label{rho_table}
\end{table}

\begin{table}
\centering
\caption{Table summarising the mean area of sky over which we could have detected each variable source. The mean area was used to calculate the sky surface densities of the sources shown in the last column. The top four sources are the short duration variables (26 minutes) and the quoted surface densities are upper limits, because we conclude the variability is from non-astrophysical origin. The bottom two represent the long duration variables (404 days) and the quoted surface densities represent the true on-sky rate.}
\begin{tabular}{c|c|c|c|c}
\hline  Source name & Mean Flux  & Area & $\rho$    \\
 &  (Jy)  &  (deg$^{2}$) &  (deg$^{-2}$)   \\
\hline
MWA J0852-2048 &   11.2 & 859 $\pm$ 97    &   $<$1.16$\times10^{-3}$  \\
MWA J0922-1428 &   ~6.1 & 860 $\pm$ 97      &    $<$2.32$\times10^{-3}$ \\
MWA J1005-2148 & 19.6   & 1065 $\pm$ 92  &    $<$2.81$\times10^{-3}$  \\
MWA J0901-2556 &   27.6 & 1175 $\pm$ 87  &    $<$3.40$\times10^{-3}$   \\
\hline
MWA J0837-1950 &   ~6.0 & 894 $\pm$ 96  &    1.11$\times10^{-3}$   \\
MWA J0901-2556 &   28.8 & 1135 $\pm$ 90  &    1.76$\times10^{-3}$   \\
\hline
\end{tabular}
\label{rho_var_table}
\end{table}

As no transients were detected, upper limits on the surface density can be placed using Poisson statistics via:

\begin{equation}
P = e^{-\rho A},
\label{snap_eq}
\end{equation}

\noindent where $\rho$ is the surface density of sources per square degree. The equivalent solid angle is given by $A$, which is found by multiplying the number of epochs $N_{e} -1 = 50$ by the area of sky surveyed per-epoch $\Omega$. The area of sky per-epoch is a function of primary beam correction; we therefore use the different  sky areas (see Table \ref{rho_table}) to calculate the surface densities. We assume that there is an equal chance of detecting a transient in each of the 51 images. The Poisson 2$\sigma$ confidence interval can be defined as $P = 0.05$ for zero detections of transients. Substituting all of the above into Equation \ref{snap_eq} and solving for~$\rho$ yields the upper limits summarised in Table~\ref{rho_table}. At the most sensitive region of the primary beam above a flux level of 5.5 Jy, the survey was sensitive to an average area of sky 802~deg$^{2}$: which gives an upper limit on the surface density of $\rho<7.5 \times 10^{-5}$ deg$^{-2}$, at this flux level.

For the short duration variable sources we used the sensitivity maps (see Section \ref{science_quality _verification}) to calculate the mean area over which each variable source could have been detected. The mean area was calculated using the modulation index $m$ of each source, which dictated the minimum signal-to-noise ratio ($1/m$) that each source would require, so that the variations could have been detected. This average area was then used to calculate the upper limits on the surface density of short duration variable sources quoted in Table~\ref{rho_var_table}. We consider these surface densities to be upper limits because we conclude that the variability is from a non-astrophysical origin. For the long duration variable sources we calculate the surface densities using the same method. The values are also quoted in Table ~\ref{rho_var_table}. We regard the long duration variables as being astrophysical in nature, therefore these values are surface densities, not upper limits. 

\begin{figure}
\centering
\includegraphics[scale=0.65]{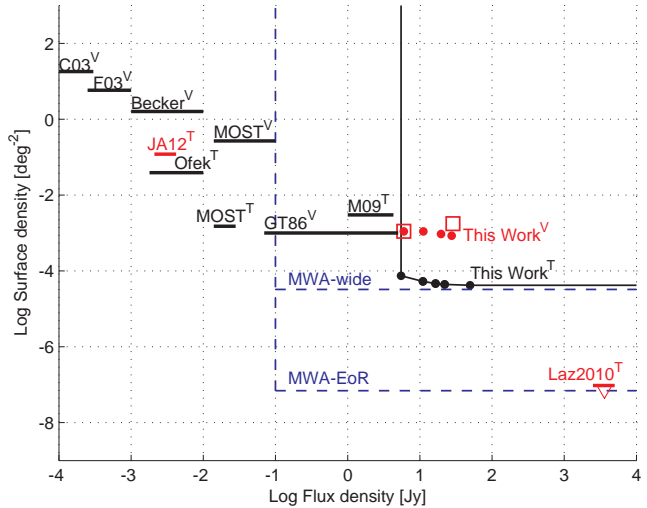}
\caption{Log surface density (deg$^{-2}$) against flux density (Jy) for a selection of surveys that have either detected transient or variable sources, or placed upper limits. Surveys for transients are labelled ``$T$" while surveys for variables are labelled ``$V$". See text for descriptions of surveys plotted. The low frequency surveys are highlighted in red and labelled JA12$^{T}$ (\citealt{Jaeger2012}) and Laz2010$^{T}$ (\citealt{Lazio2011}). Note that the \citealt{Lazio2011} is an upper limit on the surface density derived from non-detections. The upper limits on the surface density of transients from this survey are indicated with the thin black  lines and dots respectively (labelled - This Work$^{T}$). The red circles and squares, labelled This Work$^{V}$, show the upper limit on the surface densities of the four short duration variable sources (circles) and the surface densities of the two long duration variables (squares), also see Table~\ref{rho_var_table}. MWA-wide and MWA-EoR denote limits that could be placed with upcoming extra-galactic MWA surveys.} 
\label{LogN_LogS}
\end{figure}

In Figure~\ref{LogN_LogS} we compare the upper limits derived for transient sources from this survey with a selection of other surveys for transients and variables. We have selected these surveys because they predominantely report detections. The horizontal thin black line (labelled This Work$^{T}$) shows the upper limits on the surface density of transients derived above; the vertical line shows the detection limit of 5.5~Jy. 

Figure \ref{LogN_LogS} also shows the upper limits on the surface densities of the short duration variable sources and the long duration surface densities (both labelled This Work$^{V}$). Note that the four short duration variable sources (red circles) have modulation indices below 20\%. Some of the other surveys (discussed below) shown in Figure \ref{LogN_LogS} define a variable source as having $m>50\%$, or use different statistics altogether.

The other surveys shown in Figure \ref{LogN_LogS} are as follows. \citet{Carilli2003} (labelled C03$^{V}$) observed the Lockman Hole region to a sensitivity limit of $100\mu$Jy at 1.4~GHz with five epochs over a period of 17 months. They report that 2\% of the sources within their sample were highly variable (defined as $|\Delta S| \ge 50\%$) and they quote a surface density of $\rho = 18$~deg$^{-2}$. No transient sources were reported. At 4.8 GHz \citet{Becker10} (labelled Becker$^{V}$) surveyed the Galactic plane and found a surface density of variables $\rho = 1.6$~deg$^{-2}$, above a sensitivity 1~mJy. At 5~GHz \citet{Ofek2011} (labelled Ofek$^{T}$) searched for both transients and variables on a multitude of timescales (days to years) to a sensitivity limit of $\sim100 \mu$Jy: one transient candidate was reported which gave a surface density $\rho = 0.039$ deg$^{-2}$.

 \citet{Mat2009} (labelled - M09$^{T}$) summarise the detections of nine candidate transient sources at 1.4~GHz which have flux densities greater than 1~Jy, with typical timescales of minutes to days. Frail et al. (2003) labelled F03$^{V}$ found four highly variable radio transient sources from follow-up observations of GRBs at 5 and 8.5 GHz. \citet{Keith2011, Keith2011b} report the detection of two transient sources at 843~MHz from a survey with a two-epoch equivalent area of 2776~deg$^2$, giving an overall snapshot rate of $\rho = 1.5\times10^{-3}$~deg$^{-2}$ for sources above 14~mJy (labelled MOST$^{T}$). 55 variable radio sources were reported on timescales of months to years giving a rate of $\rho=$0.268~deg$^{-2}$ (labelled MOST$^{V}$). 

For further details of all transient and variable radio surveys, including papers not discussed above which report upper limits from non-detections, see \cite{FenderBell} and \cite{VASTPaper}. Also see \cite{Bower2007} and \cite{Frail2012} for further discussion of the rates of radio transients at GHz frequencies.  

To highlight how poorly studied the low frequency transient and variable sky is, we have coloured the only other blind surveys below 500 MHz in red. These surveys are those by \cite{Lazio2011} and \cite{Jaeger2012} (which are labelled JA12$^{T}$ and Laz2010$^{T}$, respectively, in Figure \ref{LogN_LogS}). The survey conducted by \cite{Lazio2011} at 74~MHz sampled a similar timescale to this survey. Around 105 hours of data were obtained with the LWDA and were searched for transient sources. A integration time of two to five minutes was used to collect just under 30,000 snapshot observations. 

\cite{Lazio2011} report no transient radio sources at 74 MHz, which places an upper limit on the event rate of $10^{-2}$ events yr$^{-1}$ deg$^{-2}$. This rate approximately converts to a surface density of $\rho<9.5\times10^{-8}$ deg$^{-2}$ (assuming that there are $\sim$10$^{5}$  five minute snapshot observations in one year), for sources with fluxes greater than 2.5 kJy (5$\sigma$). Our survey probes a flux limit that is three orders of magnitude deeper than \cite{Lazio2011} with $\sim$five times less solid angle. 

\cite{Jaeger2012} search for both variability in known sources and unique transient type sources (at 325 MHz). One radio transient is reported and speculated to be of stellar origin, perhaps via a coherent radio emission mechanism. The surface density of $\rho=0.12$ deg$^{-2}$ derived for this source is indicated on Figure \ref{LogN_LogS}. Of the 950 sources reported by \cite{Jaeger2012}, a moderate number with SNR~$>$~20 showed a modulation index of $\sim$10\%, whilst sources with SNR~$>$~50 showed a modulation index of below 5\% (with timescales of days to months). 

The survey conducted by \cite{Jaeger2012} is significant because it predicts that in a field of view of~1430~deg$^{2}$ (i.e. one MWA pointing) at a detection threshold of $>$2.1 mJy with $\rho=0.12$ deg$^{-2}$, that 172 transients of this type would be detected. The full MWA instrument will achieve a typical RMS per-snapshot of $\sim$10 mJy. Assuming a crude isotropic scaling of $S_{\nu}^{-3/2}$ and scaling the rate reported by \cite{Jaeger2012} to a sensitivity of 50~mJy (i.e. a 5$\sigma$ detection with the MWA) we would predict the detection of $<$2 transients. Scaling to a sensitivity of 5.5~Jy (which is the deepest detection limit of this survey), we would predict $< 1.0 \times 10^{-3}$ transients over the 1430~deg$^{2}$ field of view. We place an upper limit of $<6 \times 10^{-2}$ transients over the 1430~deg$^{2}$ field of view, which is consistent with the work of \cite{Jaeger2012}. 
 
In Figure \ref{LogN_LogS} we include possible limits that could be placed with upcoming MWA surveys. MWA-wide denotes a survey that will sweep the Southern Hemisphere on a monthly cadence. We assume in the upper limits shown in Figure \ref{LogN_LogS} (assuming that no transient detections are made) that $\sim$18,500 deg$^{2}$ will be re-observed up to five times down to a sensitivity limit 100 mJy. The survey will provide multiple measurements of tens of thousands of radio sources (on monthly cadences) and provide a census of transient and variable activity in the Southern sky. MWA-EoR denotes a commensal transient program that will utilise (potentially) over a thousands of hours of telescope time to detect the epoch of reionisation. In the upper limits shown in Figure \ref{LogN_LogS} we assume that 1000 hours of time will be spend observing an area of sky 1430 deg$^{2}$ (i.e a single pointing). This survey will probe characteristic timescales of $\sim$2 minutes and produce high cadence light curves for all sources within the field.     

%%%%%%%%%%%%%%%%%%%%%%%%%%
\section{Conclusion}
%%%%%%%%%%%%%%%%%%%%%%%%%%
In this paper we have presented the results of a search for low frequency transients and variables. Four short duration variable sources are found that displayed variability on timescales of minutes to days. We discuss the plausible physical scenarios for this variability and we rule out intrinsic variability and refractive, diffractive and inter-planetary scintillation as possible causes. We suggest that ionospheric or instrumental effects could explain the low levels of variability observed. We find two sources that show significant variability on the timescale of 404 days. We conclude that the mechanism for this variability is either refractive scintillation or intrinsic to the source, or both. No transients were detected and we place constraints on the prevalence of such events on timescales of minutes to years.

One of the aims of this work was to demonstrate and test the functionality of the VAST pipeline. It was also to verify the science quality of the MWA system for transient radio science in future operations. By characterising the variability of a sample of 105 sources we have demonstrated the stability of the instrument. Due to this stability we are in a regime where we could detect highly variable and transient sources in future work.   

Through this work two correlations were found which impacted our ability to use variability search statistics (such as $\chi^{2}$). A correlation was found between the off-source pixels within the individual XX and YY polarisation images (same observation), and also the individual images/observations spaced close together in time. It is probable that these correlations will also be prominent in the full MWA system. Techniques will be needed to mitigate correlated images, such as image subtraction, or more advanced statistical tests.

With the 128 tile system, we anticipate the regular detection of up to 10$^{5}$ radio sources on a variety of timescales. We will therefore be capable of rigorously quantifying the variability of low frequency radio sources in the Southern Hemisphere. The MWA is also sensitive to a broad frequency range (80$-$300 MHz), in this paper we only focus on a single 30.72 MHz band centred at 154 MHz. Future studies will focus on a broader frequency range, which will provide greater spectral information about detected sources. The full 128 tile MWA system will achieve sensitivities of $\sim$10 mJy beam$^{-1}$, over thousands of square degrees, with a resolution $\sim$ $2^{\prime}$ (see \citealt{Tingay2012}). With an increased throughput of images from the MWA into the VAST pipeline, we can start to comprehensively map out low frequency parameter space for variable and transient behaviour.  
  
 %%%%%%%%%%%%%%%%%%%%%%%%%% 
\section{Acknowledgements}
%%%%%%%%%%%%%%%%%%%%%%%%%%
This scientific work makes use of the Murchison Radio-astronomy Observatory, operated by CSIRO. We acknowledge the Wajarri Yamatji people as the traditional owners of the Observatory site. Support for the MWA comes from the U.S. National Science Foundation (grants AST-0457585, PHY-0835713, CAREER-0847753, and AST-0908884), the Australian Research Council (LIEF grants LE0775621 and LE0882938), the U.S. Air Force Office of Scientific Research (grant FA9550-0510247). This work was supported by the Centre for All-sky Astrophysics (CAASTRO), an Australian Research Council Centre of Excellence (grant CE110001020) and through the Science Leveraging Fund of the New South Wales Department of Trade and Investment. Support is also provided by the Smithsonian Astrophysical Observatory, the MIT School of Science, the Raman Research Institute, the Australian National University, and the Victoria University of Wellington (via grant MED-E1799 from the New Zealand Ministry of Economic Development and an IBM Shared University Research Grant). The Australian Federal government provides additional support via the National Collaborative Research Infrastructure Strategy, Education Investment Fund, and the Australia India Strategic Research Fund, and Astronomy Australia Limited, under contract to Curtin University. We acknowledge the iVEC Petabyte Data Store, the Initiative in Innovative Computing and the CUDA Center for Excellence sponsored by NVIDIA at Harvard University, and the International Centre for Radio Astronomy Research (ICRAR), a Joint Venture of Curtin University and The University of Western Australia, funded by the Western Australian State government.

\appendix

\label{lastpage}

\end{document}